%% file: ms.tex
\documentclass[runningheads]{llncs}

\input{preambular}

\newcommand{\proofstatement}{A proof of Theorem \ref{thm:detect} is given in
  \url{https://arxiv.org/abs/2305.18198}.}

\title{Model Checking Race-freedom When ``Sequential Consistency
  for Data-race-free Programs'' is Guaranteed\thanks{Version 2 of 20
    July 2023}}

\titlerunning{Model Checking and Sequential Consistency for
  Data-race-free Programs}

\author{
  Wenhao Wu\inst{1}\Envelope\;\orcidID{0000-0002-9087-4240} \and
  Jan H\"uckelheim\inst{2}\orcidID{0000-0003-3479-6361} \and
  Paul D.\ Hovland\inst{2}\orcidID{0000-0002-0907-2567} \and\\
  Ziqing Luo\inst{1}\orcidID{0000-0001-6557-3692} \and
  Stephen F.\ Siegel\inst{1}\orcidID{0000-0001-9359-3332}
}

\authorrunning{W.\ Wu et al.}

\institute{
  University of Delaware, Newark DE 19716, USA\\
  \email{\{wuwenhao,ziqing,siegel\}@udel.edu} \and
  Argonne National Laboratory, Lemont IL 60439, USA\\
  \email{\{jhueckelheim,hovland\}@anl.gov}
}

\date{July 20, 2023}

\begin{document}

\maketitle

\begin{abstract}
  Many parallel programming models guarantee that if all sequentially
  consistent (SC) executions of a program are free of data races, then
  all executions of the program will appear to be sequentially
  consistent.  This greatly simplifies reasoning about the program,
  but leaves open the question of how to verify that all SC executions
  are race-free.  In this paper, we show that with a few simple
  modifications, model checking can be an effective tool for verifying
  race-freedom.  We explore this technique on a suite of C programs
  parallelized with OpenMP.
  \keywords{data race \and model checking \and OpenMP}
\end{abstract}

\section{Introduction}
\label{sec:intro}
\input{sec1-intro}

\section{Theory}
\label{sec:model}
\input{sec2-theory}

\section{Implementation and Evaluation}
\label{sec:impl}
\input{sec3-impl}
\input{sec4-eval}

\section{Related Work}
\label{sec:related}
\input{sec5-related}

\section{Conclusion}
\label{sec:conclude}
\input{sec6-conclude}

\subsubsection*{Acknowledgements.}
This material is based upon work by the RAPIDS Institute, supported by
the U.S.\ Department of Energy, Office of Science, Office of Advanced
Scientific Computing Research, Scientific Discovery through Advanced
Computing (SciDAC) program, under contract DE-AC02-06CH11357 and award
DE-SC0021162.  Support was also provided by U.S.\ National Science
Foundation awards CCF-1955852 and CCF-2019309.

\bibliographystyle{splncs04}

\input{ms.bbl}
\appendix
\clearpage
\input{appendix}
\end{document}

%% file: preambular.tex
\usepackage[T1]{fontenc}
\usepackage{lmodern}
\usepackage{graphicx}

\usepackage[bookmarks,unicode,colorlinks=true,breaklinks=true]{hyperref}
\usepackage{xurl}
\hypersetup{breaklinks=true}

\usepackage{bbding}  % for the envelope symbol
\usepackage{amsmath,amsfonts}
\usepackage{listings} % Pretty-printed code with syntax highlighting
\usepackage{color}
\usepackage{cite}
\usepackage[table,xcdraw]{xcolor}
\usepackage{adjustbox}

\lstdefinelanguage{civl}{
  morekeywords=[1]{$assume, $assert, $mem, $mem_disjoint, $mem_empty, $read_set_pop, $read_set_push, $write_set_pop, $write_set_push, $local_start, $local_end, $parfor, $input, $output},
  morekeywords=[2]{int, void, for, double, if, else, while},
  morekeywords=[3]{pragma, section, sections, shared, private, reduction, atomic, seq_cst, critical, parallel, omp, schedule, dynamic, static, guided, master, single, firstprivate, lastprivate, threadprivate, ordered, omp_set_num_threads, omp_get_num_threads, omp_get_thread_num, omp_init_lock, omp_destroy_lock, omp_set_lock, omp_unset_lock, omp_test_lock, omp_get_wtime, num_threads, barrier, read, write},
  alsoletter={$},
  sensitive=false,
  morecomment=[l]{//},
  morecomment=[s]{/*}{*/},
  keywordstyle=[1]\ttfamily\color[rgb]{0,.3,.7},
  keywordstyle=[2]\ttfamily\color[rgb]{0.133,0.545,0.133},
  keywordstyle=[3]\ttfamily\color[rgb]{0.545,0.133,0.133},
  commentstyle=\color[rgb]{0.5,0.5,0.5},
  stringstyle={\color[rgb]{0.75,0.49,0.07}},
  basicstyle=\ttfamily,
  columns=fullflexible,
  keepspaces=true,
  upquote=true,
  breaklines=true,
  breakatwhitespace=true,
  numbers=left,
  frame=L,
  xleftmargin=20pt
}

\lstdefinelanguage{civlsmall}{
  morekeywords=[1]{$assume, $assert, $mem, $mem_disjoint, $mem_empty, $read_set_pop, $read_set_push, $write_set_pop, $write_set_push, $local_start, $local_end, $parfor, $input, $output},
  morekeywords=[2]{int, void, for, double, if, else, while},
  morekeywords=[3]{pragma, section, sections, shared, private, reduction, atomic, seq_cst, critical, parallel, omp, schedule, dynamic, static, guided, master, single, firstprivate, lastprivate, threadprivate, ordered, omp_set_num_threads, omp_get_num_threads, omp_get_thread_num, omp_init_lock, omp_destroy_lock, omp_set_lock, omp_unset_lock, omp_test_lock, omp_get_wtime, num_threads, barrier, read, write},
  basicstyle=\ttfamily\scriptsize,
  columns=fullflexible,
  keepspaces=true,
  alsoletter={$},
  sensitive=false,
  morecomment=[l]{//},
  morecomment=[s]{/*}{*/},
  keywordstyle=[1]\ttfamily\color[rgb]{0,.3,.7},
  keywordstyle=[2]\ttfamily\color[rgb]{0.133,0.545,0.133},
  keywordstyle=[3]\ttfamily\color[rgb]{0.545,0.133,0.133},
  commentstyle=\color[rgb]{0.5,0.5,0.5},
  stringstyle={\color[rgb]{0.75,0.49,0.07}},
  upquote=true,
  breaklines=true,
  breakatwhitespace=true,
  numbers=left,
  frame=L,
  xleftmargin=20pt
}

\spnewtheorem{definition}{Definition}{\bfseries}{\rmfamily}

\graphicspath{{../fig/}}
\makeatletter
\RequirePackage[bookmarks,unicode,colorlinks=true]{hyperref}%
   \def\@citecolor{blue}%
   \def\@urlcolor{blue}%
   \def\@linkcolor{blue}%

\def\orcidID#1{\smash{\href{http://orcid.org/#1}{\protect\raisebox{-1.25pt}{\protect\includegraphics{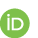}}}}}
\makeatother

\newcommand{\code}[1]{\texttt{#1}}
\newcommand{\U}{\texttt{\char`\_}}

\newcommand{\mysf}[1]{\textrm{\textup{\textsf{#1}}}}
\newcommand{\myemph}[1]{\emph{#1}}
\newcommand{\tr}{\mysf{tr}}
\newcommand{\HB}{\mysf{HB}}
\newcommand{\DR}{\mysf{DR}}
\newcommand{\N}{\mathbb{N}}
\newcommand{\State}{\mysf{State}}
\newcommand{\state}{\mysf{state}}
\newcommand{\LocalState}{\mysf{Local}}
\newcommand{\LockState}{\mysf{LockState}}
\newcommand{\LockingState}{\mysf{Acquire}}
\newcommand{\UnlockingState}{\mysf{Release}}
\newcommand{\BarrierState}{\mysf{Barrier}}

\newcommand{\NSyncState}{\mysf{Nsync}}
\newcommand{\TermState}{\mysf{Term}}

\newcommand{\nxt}{\mysf{next}}

\newcommand{\stmts}{\mysf{stmts}}
\newcommand{\update}{\mysf{update}}

\newcommand{\Shared}{\mysf{Shared}}
\newcommand{\Event}{\mysf{Event}}
\newcommand{\Lock}{\mysf{Lock}}
\newcommand{\lock}{\mysf{acquire}}
\newcommand{\lockOf}{\mysf{lock}}
\newcommand{\unlock}{\mysf{release}}

\newcommand{\barrierExit}{\mysf{exit}}

\newcommand{\Stmt}{\mysf{Stmt}}

\newcommand{\ra}{\rightarrow}
\newcommand{\conflict}{\mysf{conflict}}
\newcommand{\tid}{\mysf{tid}}
\newcommand{\epoch}{\mysf{epoch}}

\newcommand{\enabled}{\mysf{enabled}}

\newcommand{\execute}{\mysf{execute}}
\newcommand{\TID}{\mysf{TID}}
\newcommand{\ample}{\mysf{ample}}

\newcommand{\stategraph}{state graph}
\newcommand{\graph}{G}
\newcommand{\ma}{\mathbf{a}}

%% file: sec1-intro.tex
% Introduction

% ISO/IEC 9899:2017, Sec. 5.1.2.4.

% NOTE 18 It can be shown that programs that correctly use simple
% mutexes and memory_order_seq_cst operations to prevent all data races,
% and use no other synchronization operations, behave as though the
% operations executed by their constituent threads were simply
% interleaved, with each value computation of an object being the last
% value stored in that interleaving. This is normally referred to as
% “sequential consistency”. However, this applies only to data-race-free
% programs, and data-race-free programs cannot observe most program
% transformations that do not change single-threaded program
% semantics. In fact, most single-threaded program transformations
% continue to be allowed, since any program that behaves differently as
% a result must contain undefined behavior.

%Boehm, https://open-std.org/jtc1/sc22/wg21/docs/papers/2007/n2392.html.

% From a pure correctness perspective, condition variable notification
% can be modelled as a no-op, and a condition variable wait as a an
% unlock() followed by a lock() operation. Hence the results here also
% apply to programs with condition variables.

% http://rsim.cs.illinois.edu/Pubs/08PLDI.pdf

Every multithreaded programming language requires a memory model to
specify the values a thread may obtain when reading a variable.  The
simplest such model is \emph{sequential consistency}
\cite{lamport:1979:sequential-consistency}.  In this model, an
execution is an interleaved sequence of the execution steps from each
thread. The value read at any point is the last value that was written
to the variable in this sequence.

There is no known efficient way to implement a full sequentially
consistent model. One reason for this is that many standard compiler
optimizations are invalid under this model. Because of this, most
multithreaded programming languages (including language extensions)
impose a requirement that programs do not have \emph{data races}. A
data race occurs when two threads access the same variable without
appropriate synchronization, and at least one access is a write. (The
notion of appropriate synchronization depends on the specific
language.) For data race-free programs, most standard compiler
optimizations remain valid. The Pthreads library is a typical example,
in that programs with data races have no defined behavior, but
race-free programs are guaranteed to behave in a sequentially
consistent manner \cite{posix}.

Modern languages use more complex ``relaxed'' memory models. In this
model, an execution is not a single sequence, but a set of events
together with various relations on those events. These
relations---e.g., \emph{sequenced before}, \emph{modification order},
\emph{synchronizes with}, \emph{dependency-ordered before},
\emph{happens before} \cite{iso:2018:c}---must satisfy a set of
complex constraints spelled out in the language specification. The
complexity of these models is such that only the most sophisticated
users can be expected to understand and apply them correctly.
Fortunately, these models usually provide an escape, in the form of a
substantial and useful language subset which is guaranteed to behave
sequentially consistently, as long as the program is race-free.
Examples include Java \cite{manson-pugh-adve:2005:java}, C and C++
since their 2011 versions (see \cite{boehm-adve:2008:foundations} and
\cite[\S5.1.2.4 Note 19]{iso:2018:c}), and OpenMP
\cite[\S1.4.6]{omp:2021:spec52}.

The ``guarantee'' mentioned above actually consists of two parts: (1)
all executions of data race-free programs in the language subset are
sequentially consistent, and (2) if a program in the language subset
has a data race, then it has a sequentially consistent execution with
a data race \cite{boehm-adve:2008:foundations}.  Putting these
together, we have, for any program $P$ in the language subset:
\begin{center}
  \parbox{.8\textwidth}{(SC4DRF) \em If all sequentially consistent
    executions of $P$ are data race-free, then all executions of $P$
    are sequentially consistent.}
\end{center}    
The consequence of this is that the programmer need only understand
sequentially consistent semantics, both when trying to ensure $P$ is
race-free, and when reasoning about other aspects of the correctness
of $P$.  This approach provides an effective compromise between
usability and efficient implementation.

Still, it is the programmer's responsibility to ensure that all
sequentially consistent executions of the program are race-free.
Unfortunately, this problem is undecidable
\cite{bernstein:1966:analysis-parallel}, so no completely algorithmic
solution exists.  As a practical matter, detecting and eliminating
races is considered one of the most challenging aspects of parallel
program development.  One source of difficulty is that compilers may
``miscompile'' racy programs, i.e., translate them in unintuitive,
non-semantics-preserving ways \cite{boehm:2011:miscompile}.  After
all, if the source program has a race, the language standard imposes
no constraints, so any output from the compiler is technically
correct.

Researchers have explored various techniques for race checking.
Dynamic analysis tools (e.g., \cite{gu-mellor-crummey:2018:romp}) have
experienced the most uptake. These techniques can analyze a single
execution precisely, and report whether a race occurred, and sometimes
can draw conclusions about closely related executions. But the
behavior of many concurrent programs depends on the program input, or
on specific thread interleavings, and dynamic techniques cannot
explore all possible behaviors. Moreover, dynamic techniques
necessarily analyze the behavior of the executable code that results
from compilation. As explained above, racy programs may be
miscompiled, even possibly removing the race, in which case a
dynamic analysis is of limited use.

Approaches based on static analysis, in contrast, have the potential
to verify race-freedom. This is extremely challenging, though some
promising research prototypes have been developed (e.g.,
\cite{bora2021llov}). The most significant limitation is imprecision:
a tool may report that race-free code has a possible race--- a ``false
alarm''.  Some static approaches are also not sound, i.e., they may
fail to detect a race in a racy program; like dynamic tools, these
approaches are used more as bug hunters than verifiers.

Finite-state model checking \cite{clarke-etal:2018:mc2} offers an
interesting compromise. This approach requires a finite-state model of
the program, which is usually achieved by placing small bounds on the
number of threads, the size of inputs, or other program parameters.
The reachable states of the model can be explored through explicit
enumeration or other means.  This can be used to implement a sound and
precise race analysis of the model.  If a race is found, detailed
information can be produced, such as a program trace highlighting the
two conflicting memory accesses. Of course, if the analysis concludes
the model is race-free, it is still possible that a race exists for
larger parameter values. In this case, one can increase those values
and re-run the analysis until time or computational resources are
exhausted. If one accepts the ``small scope hypothesis''---the claim
that most defects manifest in small configurations of a system---then
model checking can at least provide strong evidence for the absence of
data races. In any case, the results provide specific information on
the scope that is guaranteed to be race-free, which can be used to
guide testing or further analysis.

The main limitation of model checking is state explosion, and one of
the most effective techniques for limiting state explosion is
\emph{partial order reduction} (POR) \cite{godefroid:1996:por-book}. A
typical POR technique is based on the following observation: from a
state $s$ at which a thread $t$ is at a ``local'' statement---i.e.,
one which commutes with all statements from other threads---then it is
often not necessary to explore all enabled transitions from $s$;
instead, the search can explore only the enabled transitions from $t$.
Usually local statements are those that access only thread-local
variables. But if the program is known to be race-free, shared
variable accesses can also be considered ``local'' for POR. This is
the essential observation at the heart of recent work on POR in the
verification of Pthreads programs \cite{schemmel-etal:2020:symb-por}.

In this paper, we explore a new model checking technique that can be
used to verify race-freedom, as well as other correctness properties,
for programs in which threads synchronize through locks and barriers.
The approach requires two simple modifications to the standard state
reachability algorithm.  First, each thread maintains a history of the
memory locations accessed since its last synchronization operation.
These sets are examined for races and emptied at specific
synchronization points.  Second, a novel POR is used in which only lock
(release and acquire) operations are considered non-local.  In Section
\ref{sec:model}, we present a precise mathematical formulation of the
technique and a theorem that it has the claimed properties, including
that it is sound and precise for verification of race-freedom of
finite-state models.

Using the CIVL symbolic execution and model checking platform
\cite{siegel-etal:2015:civl_sc}, we have implemented a prototype tool,
based on the new technique, for verifying race-freedom in C/OpenMP
programs. OpenMP is an increasingly popular directive-based language
for writing multithreaded programs in C, C++, or Fortran. A large
sub-language of OpenMP has the SC4DRF guarantee.\footnote{Any OpenMP
  program that does not use non-sequentially consistent atomic
  directives, \code{omp\U{}test\U{}lock}, or
  \code{omp\U{}test\U{}nest\U{}lock} \cite[\S1.4.6]{omp:2021:spec52}}
While the theoretical model deals with locks and barriers, it can be
applied to many OpenMP constructs that can be modeled using those
primitives, such as atomic operations and critical sections.  This is
explained in Section \ref{sec:impl}, along with the results of some
experiments applying our tool to a suite of C/OpenMP programs.  In
Section \ref{sec:related}, we discuss related work and Section
\ref{sec:conclude} concludes.

% The OpenMP Specification states: ``OpenMP programs that: [d]o not use
% non-sequentially consistent atomic directives; [d]o not rely on the
% accuracy of a false result from \verb!omp_test_lock! and
% \verb!omp_test_nest_lock!; and [c]orrectly avoid data races\ldots
% behave as though operations on shared variables were simply
% interleaved in an order consistent with the order in which they are
% performed by each thread.'' \cite[\S1.4.6]{omp:2021:spec52}

% Suppose there is a method to determine when any multithreaded
% program P is data-race free for all inputs.
% Given any program T, let P be the program that takes as input n>0
% and does this:
% thread 1: if (T terminates in <= n steps) x=1;
% thread 2: x=2
% if P has a data race then there is some n s.t. T terminates.
% if P does not have a data race then T does not terminate.

%% file: sec2-theory.tex
We begin with a simple mathematical model of a multithreaded program
that uses locks and barriers for synchronization.  

\begin{definition}
  Let $\TID$ be a finite set of positive integers.  A
  \myemph{multithreaded program with thread ID set $\TID$} comprises
  \begin{enumerate}
  \item a set $\Lock$ of \myemph{locks}
  \item a set $\Shared$ of \myemph{shared states}
  \item for each $i\in\TID$:
    \begin{enumerate}
    \item a set $\LocalState_i$, the \myemph{local states of thread
        $i$}, which is the union of five disjoint subsets,
      $\LockingState_i$, $\UnlockingState_i$, $\BarrierState_i$,
      $\NSyncState_i$, and $\TermState_i$
    \item a set $\Stmt_i$ of \myemph{statements}, which includes the
      \myemph{lock statements} $\lock_i(l)$ and $\unlock_i(l)$ (for
      $l\in\Lock$), and the \emph{barrier-exit} statement
      $\barrierExit_i$; all others statements are known as
      \myemph{nsync (non-synchronization) statements}
    \item for each
      $\sigma\in\LockingState_i\cup\UnlockingState_i\cup\BarrierState_i$,
      a local state $\nxt(\sigma)\in\LocalState_i$
    \item for each
      $\sigma\in\LockingState_i\cup\UnlockingState_i$,
      a lock $\lockOf(\sigma)\in\Lock$
    \item for each $\sigma\in\NSyncState_i$, a nonempty set
      $\stmts(\sigma)\subseteq\Stmt_i$ of nsync statements
      and function
      \[
        \update(\sigma)\colon \stmts(\sigma)\times\Shared\ra
        \LocalState_i\times\Shared.
      \]
    \end{enumerate}
  \end{enumerate}
  All of the sets $\LocalState_i$ and $\Stmt_i$ ($i\in\TID$) are
  pairwise disjoint. \qed
\end{definition}

Each thread has a unique thread ID number, an element of $\TID$.  A
local state for thread $i$ encodes the values of all thread-local
variables, including the program counter.  A shared state encodes the
values of all shared variables.  (Locks are not considered shared
variables.)  A thread at an \emph{acquire} state $\sigma$ is
attempting to acquire the lock $\lockOf(\sigma)$.  At a \emph{release}
state, the thread is about to release a lock.  At a \emph{barrier}
state, a thread is waiting inside a barrier.  After executing one of
the three operations, each thread moves to a unique next local state.
A thread that reaches a \emph{terminal} state has terminated.  From an
\emph{nsync} state, any positive number of statements are enabled, and
each of these statements may read and update the local state of the
thread and/or the shared state.

For $i\in\TID$, the \emph{local graph} of thread $i$ is the directed
graph with nodes $\LocalState_i$ and an edge $\sigma\ra\sigma'$ if
either (i)
$\sigma\in\LockingState_i\cup\UnlockingState_i\cup\BarrierState_i$ and
$\sigma'=\nxt(\sigma)$, or (ii) $\sigma\in\NSyncState_i$ and there is
some $\zeta'\in\Shared$ such that $(\sigma',\zeta')$ is in the image
of $\update(\sigma)$. 

Fix a multithreaded program $P$ and let
\begin{align*}
  \LockState &= (\Lock\ra \{0\} \cup \TID)\\
  \State &= \left(\prod_{i\in\TID}\LocalState_i\right)
           \times \Shared \times \LockState \times 2^{\TID}.
\end{align*}
A \emph{lock state} specifies the owner of each lock.  The owner is a
thread ID, or $0$ if the lock is free.  The elements of $\State$ are
the (global) \emph{states} of $P$.  A state specifies a local state
for each thread, a shared state, a lock state, and the set of threads
that are currently blocked at a barrier.

Let $i\in\TID$ and
$L_i=\LocalState_i \times \Shared \times \LockState \times 2^{\TID}$.
Define
\begin{gather*}
  \enabled_i\colon L_i\ra 2^{\Stmt_i}
  \\
  \lambda\mapsto
  \begin{cases}
    \{\lock_i(l)\} & \text{if $\sigma\in\LockingState_i\wedge
      l=\lockOf(\sigma)\wedge \theta(l)=0$}\\
    \{\unlock_i(l)\} & \text{if
      $\sigma\in\UnlockingState_i\wedge l=\lockOf(\sigma)\wedge
      \theta(l)=i$}\\
    \{\barrierExit_i\} & \text{if $\sigma\in\BarrierState_i\wedge
      i\not\in w$}\\
    \stmts(\sigma) & \text{if $\sigma\in\NSyncState_i$}\\
    \emptyset & \text{otherwise.}
  \end{cases}
\end{gather*}
where $\lambda=(\sigma,\zeta,\theta,w)\in L_i$.  This function
returns the set of statements that are enabled in thread $i$ at a
given state.  This function does not depend on the local states of
threads other than $i$, which is why those are excluded from $L_i$.
An acquire statement is enabled if the lock is free; a release is
enabled if the calling thread owns the lock.  A barrier exit is
enabled if the thread is not currently in the barrier blocked set.

Execution of an enabled statement in thread $i$ updates
the state as follows:
\begin{gather*}
  \execute_i\colon \{(\lambda,t)\in L_i\times \Stmt_i \mid
  t\in\enabled_i(\lambda)\} \ra L_i
  \\
  (\lambda,t) \mapsto
  \begin{cases}
    (\sigma', \zeta, \theta[l\mapsto i],w')
    & \text{if $\sigma\in\LockingState_i\wedge
      t=\lock_i(l)\wedge\sigma'=\nxt(\sigma)$} \\
    (\sigma', \zeta, \theta[l\mapsto 0],w')
    & \text{if $\sigma\in\UnlockingState_i\wedge
      t=\unlock_i(l)\wedge\sigma'=\nxt(\sigma)$} \\
    (\sigma', \zeta, \theta,w')
    &\text{if $\sigma\in\BarrierState_i\wedge
      t=\barrierExit_i\wedge\sigma'=\nxt(\sigma)$}\\
    (\sigma', \zeta', \theta,w') & \text{\parbox[t]{2in}{if
        $\sigma\in\NSyncState_i\wedge t\in\stmts(\sigma)\wedge
        \update(\sigma)(t,\zeta)=(\sigma',\zeta')$}}
  \end{cases}
\end{gather*}
where $\lambda=(\sigma,\zeta,\theta,w)$ and in each case above
\[
  w' =
  \begin{cases}
    w\cup\{i\}
    & \text{if $\sigma'\in\BarrierState_i\wedge w\cup\{i\}\neq\TID$}\\
    \emptyset
    & \text{if $\sigma'\in\BarrierState_i\wedge w\cup\{i\}=\TID$}\\
    w
    & \text{otherwise.}
  \end{cases}
\]
Note a thread arriving at a barrier will have its ID added to the
barrier blocked set, unless it is the last thread to arrive, in which
case all threads are released from the barrier.

At a given state, the set of enabled statements is the union over all
threads of the enabled statements in that thread.  Execution of a
statement updates the state as above, leaving the local states of
other threads untouched:
\begin{gather*}
  \enabled\colon \State \ra 2^{\Stmt}
  \\
  s \mapsto \bigcup_{j\in\TID}\enabled_j(\xi_j, \zeta, \theta, w)
  \\
  \execute\colon \{(s,t)\in\State \times \Stmt\mid t\in\enabled(s)\}
  \ra \State
  \\
  (s, t) \mapsto \langle \xi[i\mapsto\sigma], \zeta', \theta', w' \rangle,
\end{gather*}
where $s=\langle \xi, \zeta, \theta, w \rangle \in \State$,
$t\in\enabled(s)$, $i=\tid(t)$, and\\
$\execute_{i}(\xi_{i}, \zeta, \theta, w, t) = \langle\sigma, \zeta',
\theta',w'\rangle$.

\begin{definition}
  \label{def:execution}
  A \myemph{transition} is a triple $s\stackrel{t}{\ra}s'$, where
  $s\in\State$, $t\in\enabled(s)$, and $s'=\execute(s,t)$.  An
  \myemph{execution} $\alpha$ of $P$ is a (finite or infinite) chain
  of transitions
  \( s_0\stackrel{t_1}{\ra}s_1\stackrel{t_2}{\ra}\cdots \).
  The \myemph{length} of $\alpha$, denoted $|\alpha|$, is the number
  of transitions in $\alpha$. \qed
\end{definition}
Note that an execution is completely determined by its initial
state $s_0$ and its statement sequence $t_1t_2\cdots$.

Having specified the semantics of the computational model, we now turn
to the concept of the \emph{data race}.  The traditional definition
requires the notion of ``conflicting'' accesses: two accesses to the
same memory location conflict when at least one is a write.  The
following abstracts this notion:
\begin{definition}
  \label{def:conflict}
  A symmetric binary relation \conflict{} on $\Stmt$ is a
  \myemph{conflict relation} for $P$ if the following hold for all
  $t_1,t_2\in\Stmt$:
  \begin{enumerate}
  \item if $(t_1,t_2)\in\conflict$ then $t_1$ and $t_2$ are nsync
    statements from different threads
  \item if $t_1$ and $t_2$ are nsync statements from different threads
    and $(t_1,t_2)\not\in\conflict$, then for all $s\in\State$, if
    $t_1,t_2\in\enabled(s)$ then\\[1mm]
    \phantom{.}\hspace{5em}
    \(
    \execute(\execute(s,t_1),t_2) = \execute(\execute(s,t_2),t_1).
    \) \qed
  \end{enumerate}
\end{definition}
Fix a conflict relation for $P$ for the remainder of this section.

The next ingredient in the definition of \emph{data race} is the
\emph{happens-before} relation.  This is a relation on the set of
\emph{events} generated by an execution.  An event is an element of
$\Event=\Stmt\times\N$.
\begin{definition}
  Let $\alpha = (s_0\stackrel{t_1}{\ra}s_1\stackrel{t_2}{\ra}\cdots)$
  be an execution.  The \myemph{trace of $\alpha$} is the sequence of
  events
  $\tr(\alpha)=\langle t_1,n_1\rangle\langle t_2,n_2\rangle\cdots$, of
  length $|\alpha|$, where $n_i$ is the number of $j\in[1,i]$ for
  which $\tid(t_j)=\tid(t_i)$.
  We write $[\alpha]$ for the set of events occurring in
  $\tr(\alpha)$. \qed
\end{definition}
A trace labels the statements executed by a thread with consecutive
integers starting from $1$.  Note the cardinality of $[\alpha]$ is
$|\alpha|$, as no two events in $\tr(\alpha)$ are equal.  Also,
$[\alpha]$ is invariant under transposition of two adjacent
commuting transitions from different threads.

Given an execution $\alpha$, the \myemph{happens-before relation of
  $\alpha$}, denoted $\HB(\alpha)$, is a binary relation on
$[\alpha]$.  It is the transitive closure of the union of three
relations:
\begin{enumerate}
\item the intra-thread order relation
  \[
    \{(\langle t_1,n_1\rangle,
    \langle t_2,n_2\rangle)\in[\alpha]\times[\alpha] \mid
    \tid(t_1)=\tid(t_2)\wedge n_1<n_2\}.
  \]
\item the release-acquire relation.  Say $\tr(\alpha)=e_1e_2\ldots$
  and $e_i=\langle t_i,n_i\rangle$.  Then $(e_i,e_j)$ is in the
  release-acquire relation if there is some $l\in\Lock$ such that all
  of the following hold: (i) $1\leq i<j\leq|\alpha|$, (ii) $t_i$ is a
  release statement on $l$, (iii) $t_j$ is an acquire statement on
  $l$, and (iv) whenever $i<k<j$, $t_k$ is not an acquire statement on
  $l$.
\item the barrier relation.  For any
  $e=\langle t,n\rangle\in[\alpha]$, let $i=\tid(t)$ and define
  \[
    \epoch(e)=|\{e'\in[\alpha]\mid
    \text{$e'=\langle\barrierExit_i,j\rangle$ for some $j\in[1,n]$}\}|,
  \]  
  the number of barrier exit events in thread $i$ preceding or
  including $e$.  The barrier relation is
  \[
    \{(e,e')\in[\alpha]\times[\alpha]\mid \epoch(e)<\epoch(e')\}.
  \]
\end{enumerate}
Two events ``race'' when they conflict but are not ordered by
happens-before:
\begin{definition}
  \label{def:race}
  Let $\alpha$ be an execution and $e,e'\in [\alpha]$. Say
  $e=\langle t,n\rangle$ and $e'=\langle t',n'\rangle$.  We say $e$
  and $e'$ \myemph{race in $\alpha$} if $(t,t')\in\conflict$ and
  neither $(e,e')$ nor $(e',e)$ is in $\HB(\alpha)$.  The \myemph{data
    race relation of $\alpha$} is the symmetric binary relation on
  $[\alpha]$\\
  \phantom{x}\hspace{.5in}
  \( \DR(\alpha) = \{ (e,e')\in[\alpha]\times[\alpha]\mid \text{$e$
    and $e'$ race in $\alpha$} \}.  \) \qed
\end{definition}

Now we turn to the problem of detecting data races.  Our approach is
to explore a modified state space.  The usual state space is a
directed graph with node set $\State$ and transitions for edges.  We
make two modifications.  First, we add some ``history'' to the state.
Specifically, each thread records the nsync statements it has executed
since its last lock event or barrier exit.  This set is checked
against those of other threads for conflicts, just before it is
emptied after its next lock event or barrier exit.  The second change
is a reduction: any state that has an enabled statement that is not a
lock statement will have outgoing edges from only one thread in the
modified graph.

A well-known technical challenge with partial order reduction concerns
cycles in the reduced state space.  We deal with this challenge by
assuming that $P$ comes with some additional information.
Specifically, for each $i$, we are given a set $R_i$, with
$\UnlockingState_i\cup\LockingState_i\subseteq
R_i\subseteq\LocalState_i$, satisfying: any cycle in the local graph
of thread $i$ has at least one node in $R_i$.  In general, the smaller
$R_i$, the more effective the reduction.  In many application domains,
there are no cycles in the local graphs, so one can take
$R_i=\UnlockingState_i\cup\LockingState_i$.  For example, standard
\emph{for} loops in C, in which the loop variable is incremented by a
fixed amount at each iteration, do not introduce cycles, because the
loop variable will take on a new value at each iteration.  For
\emph{while} loops, one may choose one node from the loop body to be
in $R_i$.  \emph{Goto} statements may also introduce cycles and could
require additions to $R_i$.

\begin{definition}
  \label{def:race-space}
  The \emph{race-detecting \stategraph{}} for $P$ is the pair
  $G=(V,E)$, where
  \[
    V=\State\times\Bigl(\prod_{i\in\TID}2^{\Stmt_i}\Bigr)
  \]
  and $E\subseteq V\times\Stmt\times V$ consists of all
  $(\langle s,\ma\rangle, t, \langle s',\ma'\rangle)$ such that,
  letting $\sigma_i$ be the local state of thread $i$ in $s$,
  \begin{enumerate}
  \item $s\stackrel{t}{\ra}s'$ is a transition in $P$
  \item $\forall i\in\TID$,
    \(
    \ma'_i=
    \begin{cases}
      \ma_i\cup\{t\} & \text{if $t$ is an nsync statement
        in thread $i$}\\
      \emptyset & \text{if $t=\barrierExit_0$ or
        $i=\tid(t)\wedge\sigma_i\in R_i$} \\
      \ma_i & \text{otherwise}
    \end{cases}
    \)
  \item if there is some $i\in\TID$ such that $\sigma_i\not\in R_i$
    and thread $i$ has an enabled statement at $s$, then $\tid(t)$ is
    the minimal such $i$.  \qed
  \end{enumerate}
\end{definition}
The race-detecting state graph may be thought of as a directed graph
in which the nodes are $V$ and edges are labeled by statements.  Note
that at a state where all threads are in the barrier, $\barrierExit_0$
is the only enabled statement in the race-detecting state graph, and
its execution results in emptying all the $\ma_i$.  A lock event in
thread $i$ results in emptying $\ma_i$ only.

\begin{definition}
  \label{def:detect}
  Let $P$ be a multithreaded program and $\graph=(V,E)$ the
  race-detecting \stategraph{} for $P$.
  \begin{enumerate}
  \item Let $u=\langle s,\ma\rangle\in V$ and $i\in\TID$.  We say
    \myemph{thread $i$ detects a race in $u$} if there exist
    $j\in\TID\setminus\{i\}$, $t_1\in \ma_i$, and $t_2\in\ma_j$ such
    that $(t_1,t_2)\in\conflict$.
  \item Let $e=v\stackrel{t}{\ra}v'\in E$, $i=\tid(t)$, and $\sigma'$
    the local state of thread $i$ at $v'$.  We say \myemph{$e$ detects
      a race} if $\sigma'\in R_i\cup\BarrierState_i\cup\TermState_i$
    and thread $i$ detects a race in $v'$.
    % either
    % (i) $\sigma\in R_i\setminus\LockingState_i$ and thread $i$ detects
    % a race in $v$, (ii) $\sigma'\in\LockingState_i$ and thread $i$
    % detects a race in $v'$, or (ii) $t=\barrierExit_0$ and any thread
    % detects a race in $v$.
  \item We say \myemph{$\graph$ detects a race from $u$} if $E$
    contains an edge that is reachable from $u$ and detects a race.
    % or
    % there is some $v=\langle s,\ma\rangle\in V$ that is reachable from
    % $u$, and $i\in\TID$, such that $\enabled(s)=\emptyset$ and thread
    % $i$ detects a race in $v$.
    \qed
  \end{enumerate}
\end{definition}

Definition \ref{def:detect} suggests a method for detecting data races
in a multithreaded program.  The nodes and edges of the race-detecting
state graph reachable from an initial node are explored.  (The order
in which they are explored is irrelevant.)  When an edge in thread $i$
brings thread $i$ to an $R_i$, barrier, or terminal state, the
elements of $\ma_i$ are compared with those in $\ma_j$ for all
$j\in\TID\setminus\{i\}$ to see if a conflict exists, and if so, a
data race is reported.
% When an edge from a thread at an $R_i\setminus\LockingState_i$ state
% is executed, the elements of $\ma_i$ are compared with those in
% $\ma_j$ for all $j\in\TID\setminus\{i\}$ to see if a conflict
% exists, and if so, a data race is reported.  When an edge in thread
% $i$ terminates at an $\LockingState_i$ state, a similar race check
% takes place.  When an $\barrierExit_0$ occurs, or a node with no
% outgoing edges is reached, $\ma_i$ and $\ma_j$ are compared for all
% $i, j\in\TID$ with $i\neq j$.
This approach is sound and precise in the following sense:
\begin{theorem}
  \label{thm:detect}
  Let $P$ be a multithreaded program, and $\graph=(V,E)$ the
  race-detecting \stategraph{} for $P$.  Let $s_0\in\State$ and let
  $u_0=\langle s_0, \emptyset^{\TID}\rangle\in V$.  Assume the
  set of nodes reachable from $u_0$ is finite.  Then
  \begin{enumerate}
  \item $P$ has an execution from $s_0$ with a data race if, and only
    if, $\graph$ detects a race from $u_0$.
  \item If there is a data race-free execution of $P$ from $s_0$ to some
    state $s_f$ with $\enabled(s_f)=\emptyset$ then there is a path in
    $\graph$ from $u_0$ to a node with state component $s_f$.
  \end{enumerate}
\end{theorem}
\proofstatement
% This is one of:
% A proof of Theorem \ref{thm:detect} is given in the Appendix.
% A proof of Theorem \ref{thm:detect} is given in
% \url{https://arxiv.org/abs/2023/xxxxxxxxxx}.

\begin{example}
  Consider the $2$-threaded program represented in pseudocode:
  \begin{align*}
    t_1&\colon\ \lock(l_1)\code{;\ } \code{x=1;\ }
         \unlock(l_1)\code{;}\\
    t_2&\colon\ \lock(l_2)\code{;\ } \code{x=2;\ }
         \unlock(l_2)\code{;}
  \end{align*}
  where $l_1$ and $l_2$ are distinct locks. Let
  $R_i=\UnlockingState_i\cup\LockingState_i$ ($i=1,2$).  One path in
  the race-detecting state graph $G$ executes as follows:
  \[
    \lock(l_1)\code{;\ } \code{x=1;\ } \unlock(l_1)\code{;\ }
    \lock(l_2)\code{;\ } \code{x=2;\ } \unlock(l_2)\code{;}.
  \]
  A data race occurs on this path since the two assignments conflict
  but are not ordered by happens-before. The race is not detected,
  since at each lock operation, the statement set in the other thread
  is empty.  However, there is another path
  \[
    \lock(l_1)\code{;\ } \code{x=1;\ }
    \lock(l_2)\code{;\ } \code{x=2;\ }
    \unlock(l_1)\code{;\ }
  \]
  in $G$, and on this path the race is detected at the release.
\end{example}

%% file: sec3-impl.tex
% Implementation and Evaluation

We implemented a verification tool for C/OpenMP programs using the
CIVL symbolic execution and model checking framework.  This tool can
be used to verify absence of data races within bounds on certain
program parameters, such as input sizes and the number of threads.
(Bounds are necessary so that the number of states is finite.)  The
tool accepts a C/OpenMP program and transforms it into CIVL-C, the
intermediate verification language of CIVL.  The CIVL-C program has a
state space similar to the race-detecting state graph described in
Section \ref{sec:model}.  The standard CIVL verifier, which uses model
checking and symbolic execution techniques, is applied to the
transformed code and reports whether the given program has a data
race, and, if so, provides precise information on the variable
involved in the race and an execution leading to the race.

The approach is based on the theory of Section \ref{sec:model}, but
differs in some implementation details. For example, in the
theoretical approach, a thread records the set of non-synchronization
statements executed since the thread's last synchronization operation.
%This formulation was chosen to simplify the formal model.
This data is used only to determine whether a conflict took place
between two threads.  Any type of data that can answer this question
would work equally well.  In our implementation, each thread instead
records the set of memory locations read, and the set of memory
locations modified, since the last synchronization.  A conflict occurs
if the read or write set of one thread intersects the write set of
another read.  As CIVL-C provides robust support for tracking memory
accesses, this approach is relatively straightforward to implement by
a program transformation.

In Section \ref{sec:openmp}, we summarize the basics of OpenMP.  In
Section \ref{sec:civlc}, we provide the necessary background on CIVL-C
and the primitives used in the transformation.  In Section
\ref{sec:transform}, we describe the transformation itself. In Section
\ref{sec:eval}, we report the results of experiments using this tool.

The experiments were run using CIVL revision 5815
(\url{http://civl.dev}).  All artifacts necessary to reproduce the
experiments, as well as the full results, are available at
\url{https://github.com/verified-software-lab/sc4drf.git}.

\subsection{Background on OpenMP}
% Owner: Jan
\label{sec:openmp}

OpenMP is a pragma-based language for parallelizing programs written in C, C++ and Fortran~\cite{dagum1998openmp}. OpenMP was originally designed and is still most commonly used for shared-memory parallelization on CPUs, although the language is evolving and supports an increasing number of parallelization styles and hardware targets.
We introduce here the OpenMP features that are currently supported by our implementation in CIVL. An example that uses many of these features is shown in Figure~\ref{fig:ompexample}.

The \lstinline{parallel} construct declares the following structured block as a \emph{parallel region}, which will be executed by all threads concurrently. Within such a parallel region, programmers can use \emph{worksharing} constructs that cause certain parts of the code to be executed only by a subset of threads. Perhaps most importantly, the \emph{loop worksharing construct} can be used inside a parallel region to declare a \lstinline{omp for} loop whose iterations are mapped to different threads. The mapping of iterations to threads can be controlled through the \lstinline{schedule} clause, which can take values including \lstinline{static}, \lstinline{dynamic}, \lstinline{guided} along with an integer that defines the \emph{chunk size}. If no schedule is explicitly specified, the OpenMP run time is allowed to use an arbitrary mapping. Furthermore, a structured block within a worksharing loop may be declared as \lstinline{ordered}, which will cause this block to be executed sequentially in order of the iterations of the worksharing loop.
Worksharing for non-iterative workloads is supported through the \lstinline{sections} construct, which allows the programmer to define a number of different structured blocks of code that will be executed in parallel by different threads.

Programmers may use pragmas and clauses for \lstinline{barrier}s, \lstinline{atomic} updates, and locks. OpenMP supports named \lstinline{critical} sections, allowing no more than one thread at a time to enter a critical section with that name, and unnamed critical sections that are associated with the same global mutex. OpenMP also offers \lstinline{master} and \lstinline{single} constructs that are executed only by the \emph{master thread} 
% (that is, thread 0) 
or one arbitrary thread.

Variables are shared by all threads by default. Programmers may change the default, as well as the scope of individual variables, for each parallel region using the following clauses: \lstinline{private} causes each thread to have its own variable instance, which is uninitialized at the start of the parallel region and separate from the original variable that is visible outside the parallel region. The \lstinline{firstprivate} scope declares a private variable that is initialized with the value of the original variable, whereas the \lstinline{lastprivate} scope declares a private variable that is uninitialized, but whose final value is that of the logically last worksharing loop iteration or lexically last section. The \lstinline{reduction} clause initializes each instance to the neutral element, for example $0$ for \lstinline{reduction(+)}. Instances are combined into the original variable in an implementation-defined order. 

CIVL can model OpenMP types and routines to query and control the number of threads (\lstinline{omp_set_num_threads}, \lstinline{omp_get_num_threads}), get the current thread ID (\lstinline{omp_get_thread_num}), interact with locks (\lstinline{omp_init_lock}, \lstinline{omp_destroy_lock}, \lstinline{omp_set_lock}, \lstinline{omp_unset_lock}, and obtain the current wall clock time (\lstinline{omp_get_wtime}).

\begin{figure}[t]
\begin{lstlisting}[language=civlsmall]
#pragma omp parallel shared(b) private(i) shared(u,v)
{ // parallel region: all threads will execute this
  #pragma omp sections         // sections worksharing construct
  {
    #pragma omp section        // one thread will do this...
    { b = 0; v = 0; }
    #pragma omp section        // while another thread does this...
    u = rand();
  }
  // loop worksharing construct partitions iterations by schedule.  Each thread has a
  // private copy of b; these are added back to original shared b at end of loop...
  #pragma omp for reduction(+:b) schedule(dynamic,1)
  for (i=0; i<10; i++) {
    b = b + i;
    #pragma omp atomic seq_cst // atomic update to v 
    v+=i;
    #pragma omp critical (collatz) // one thread at a time enters critical section 
    u = (u%2==0) ? u/2 : 3*u+1;
  }
}
\end{lstlisting}
\vspace{-2em}
  \caption{OpenMP Example}
  \label{fig:ompexample}
\end{figure}

\subsection{Background on CIVL-C}
\label{sec:civlc}

The CIVL framework includes a front-end for preprocessing, parsing,
and building an AST for a C program.  It also provides an API for
transforming the AST.  We used this API to build a tool which consumes
a C/OpenMP program and produces a CIVL-C ``model'' of the program.
The CIVL-C language includes most of sequential C, including
functions, recursion, pointers, structs, and dynamically allocated
memory.  It adds nested function definitions and primitives for
concurrency and verification.

In CIVL-C, a thread is created by \emph{spawning} a function:
\code{\$spawn f(...);}.  There is no special syntax for shared or
thread-local variables; any variable that is in scope for two threads
is shared.  CIVL-C uses an interleaving model of concurrency similar
to the formal model of Section \ref{sec:model}.  Simple statements,
such as assignments, execute in one atomic step.

Threads can synchronize using \emph{guarded commands}, which have the
form \mbox{\code{\$when\! (}$e$\code{)} $S$}.  The first atomic
substatement of $S$ is guaranteed to execute only from a state in
which $e$ evaluates to \emph{true}.  For example, assume thread IDs
are numbered from $0$, and a lock value of $-1$ indicates the lock is
free.  The \emph{acquire} lock operation may be implemented as
\code{\$when (l<0) l=tid;}, where \code{l} is an integer shared
variable and \code{tid} is the thread ID.  A \emph{release} is simply
\code{l=-1;}.

A convenient way to spawn a set of threads is
\mbox{\code{\$parfor (int }$i$\code{:}$d$\code{)} $S$}.
This spawns one thread for each element of the 1d-domain $d$; each
thread executes $S$ with $i$ bound to one element of the domain.  A
1d-domain is just a set of integers; e.g., if $a$ and $b$ are integer
expressions, the domain expression $a$\code{..}$b$ represents the set
$\{a,a+1,\ldots, b\}$.  The thread that invokes the \code{\$parfor} is
blocked until all of the spawned threads terminate, at which point the
spawned threads are destroyed and the original thread proceeds.

CIVL-C provides primitives to constrain the interleaving semantics of
a program.  The program state has a single atomic lock, initially
free.  At any state, if there is a thread $t$ that owns the atomic
lock, only $t$ is enabled.  When the atomic lock is free, if there is
some thread at a \code{\$local\U{}start} statement, and the first
statement following \code{\$local\U{}start} is enabled, then among
such threads, the thread with lowest ID is the only enabled thread;
that thread executes \code{\$local\U{}start} and obtains the lock.
When $t$ invokes \code{\$local\U{}end}, $t$ relinquishes the atomic
lock.  Intuitively, this specifies a block of code to be executed
atomically by one thread, and also declares that the block should be
treated as a local statement, in the sense that it is not necessary to
explore all interleavings from the state where the local is enabled.

Local blocks can also be broken up at specified points using function
\code{\$yield}.  If $t$ owns the atomic lock and calls \code{\$yield},
then $t$ relinquishes the lock and does not immediately return from
the call.  When the atomic lock is free, there is no thread at a
\code{\$local\U{}start}, a thread $t$ is in a \code{\$yield}, and the
first statement following the \code{\$yield} is enabled, then $t$ may
return from the \code{\$yield} call and re-obtain the atomic lock.
This mechanism can be used to implement the race-detecting state
graph: thread $i$ begins with \code{\$local\U{}start}, yields at each
$R_i$ node, and ends with \code{\$local\U{}end}.

CIVL's standard library provides a number of additional primitives.
For example, the concurrency library provides a barrier implementation
through a type \code{\$barrier}, and functions to initialize, destroy,
and invoke the barrier.

The \emph{mem} library provides primitives for tracking the sets of
memory locations (a variable, an element of an array, field of a
struct, etc.) read or modified through a region of code.  The type
\code{\$mem} is an abstraction representing a set of memory locations,
or \emph{mem-set}.  The state of a CIVL-C thread includes a stack of
mem-sets for writes and a stack for reads.  Both stacks are initially
empty.  The function \code{\$write\U{}set\U{}push} pushes a new empty
mem-set onto the write stack.  At any point when a memory location is
modified, the location is added to the top entry on the write stack.
Function \code{\$write\U{}set\U{}pop} pops the write stack, returning
the top mem-set.
The corresponding functions for the read stack are
\code{\$read\U{}set\U{}push} and \code{\$read\U{}set\U{}pop}.  The
library also provides various operations on mem-sets, such as
\code{\$mem\U{}disjoint}, which consumes two mem-sets and returns
\emph{true} if the intersection of the two mem-sets is empty.

% note: I changed the name from mem_no_intersect to mem_disjoint.

\subsection{Transformation for Data Race Detection}
\label{sec:transform}

\begin{figure}[t]
\begin{lstlisting}[language=civlsmall]
int nthreads = ...;
$mem reads[nthreads], writes[nthreads];
void check_conflict(int i, int j) {
  $assert($mem_disjoint(reads[i], writes[j]) && $mem_disjoint(writes[i], reads[j]) &&
          $mem_disjoint(writes[i], writes[j]));
}
void clear_all() {
  for (int i=0; i<nthreads; i++) reads[i] = writes[i] = $mem_empty();
}
void run(int tid) {
  void pop() { reads[tid]=$read_set_pop(); writes[tid]=$write_set_pop(); }
  void push() { $read_set_push(); $write_set_push(); }
  void check() {
    for (int i=0; i<nthreads; i++) { if (i==tid) continue; check_conflict(tid, i); }
  }
  // local variable declarations
  $local_start(); push(); S pop(); $local_end();
}
for (int i=0; i<nthreads; i++) reads[i] = writes[i] = $mem_empty();
$parfor (int tid:0..nthreads-1) run(tid);
check_and_clear_all();
\end{lstlisting}
  \caption{Translation of \code{\#pragma omp parallel} $S$}
  \label{fig:trans}
\end{figure}

The basic structure for the transformation of a parallel construct is
shown in Figure \ref{fig:trans}.  The user specifies on the command
line the default number of threads to use in a parallel region.  After
this, two shared arrays are allocated, one to record the read set for
each thread, and the other the write set.  Rather than updating these
arrays immediately with each read and write event, a thread updates
them only at specific points, in such a way that the shared sets are
current whenever a data race check is performed.

The auxiliary function \verb!check_conflict! asserts no read-write or
write-write conflict exists between threads $i$ and $j$.  Function
\verb!clear_all! clears all shared mem-sets.

Each thread executes function \code{run}.  A local copy of each
private variable is declared (and, for \code{firstprivate} variables,
initialized) here.  The body of this function is enclosed in a local
region.  The thread begins by pushing new entries onto its read and
write stacks.  As explained in Section \ref{sec:civlc}, this turns on
memory access tracking.  The body $S$ is transformed in several ways.
First, references to the private variable are replaced by references
to the local copy.  Other OpenMP constructs are translated as follows.

\paragraph{Lock operations.}
Several OpenMP operations are modeled using locks.  The
\code{omp\U{}set\U{}lock} and \code{omp\U{}unset\U{}lock} functions
are the obvious examples, but we also use locks to model the behavior
of atomic and critical section constructs.  In any case, a lock
acquire operation is translated to
\begin{verbatim}
  pop(); check(); $yield(); acquire(l); push();
\end{verbatim}
The thread first pops its stacks, updating its shared mem-sets.  At
this point, the shared structures are up-to-date, and the thread uses
them to check for conflicts with other threads.  This conforms with
Definition \ref{def:detect}(2), that a race check occur upon arrival
at an acquire location.  It then yields to other threads as it
attempts to acquire lock
$l$.  Once acquired, it pushes new empty entries onto its stack and
resumes tracking.  Similarly, a release statement becomes
\begin{verbatim}
  pop(); check(); $yield(); release(l); push();
\end{verbatim}
A similar sequence is inserted in any loop (e.g., a \emph{while} loop
or a \emph{for} loop not in standard form) that may create a cycle in
the local space, only without the release statement.

\paragraph{Barriers.}
An explicit or implicit barrier in $S$ becomes
\begin{small}
\begin{verbatim}
  pop(); check(); $local_end(); $barrier_call();
  if (tid==0) clear_all();
  $barrier_call(); $local_start(); push();.
\end{verbatim}
\end{small}
The CIVL-C \code{\$barrier\U{}call} function must be invoked outside
of a local region, as it may block.  Once all threads are in the
barrier, a single thread ($0$) clears all the shared mem-sets.  A
second barrier call is used to prevent other threads from racing ahead
before this clear completes.

\paragraph{Atomic and critical sections.}
An OpenMP atomic construct is modeled by introducing a global ``atomic
lock'' which is acquired before executing the atomic statement and
then released.  The acquire and release actions are then transformed
as described above.  Similarly, a lock is introduced for each critical
section name (and the anonymous critical section); this lock is
acquired before entering a critical section with that name and
released when departing.

\paragraph{Worksharing constructs.}
Upon arriving at a \code{for} construct, a thread invokes a function
that returns the set of iterations for which the thread is
responsible.  The partitioning of the iteration space among the
threads is controlled by the construct clauses and various command
line options.  If the construct specifies the distribution strategy
precisely, then the model uses only that distribution.  If the
construct does not specify the distribution, then the decisions are
based on command line options.  One option is to explore all possible
distributions. In this case, when the first thread arrives, a series
of nondeterministic choices is made to construct an arbitrary
distribution.  The verifier explores all possible choices, and
therefore all possible distributions.  This enables a complete
analysis of the loop's execution space, but at the expense of a
combinatorial explosion with the number of threads or iterations.  A
different command line option allows the user to specify a particular
default distribution strategy, such as \emph{cyclic}.  These options
give the user some control over the completeness-tractability
tradeoff.  For \code{sections}, only cyclic distribution is currently
supported, and a \code{single} construct is executed by the first
thread to arrive at the construct.

%% file: sec4-eval.tex
\subsection{Evaluation}
\label{sec:eval}

We applied our verifier to a suite comprised of benchmarks from
DataRaceBench (DRB) version 1.3.2 \cite{verma2020enhancing}
and some examples written by us that use different concurrency
patterns.  As a basis for comparison, we applied a state-of-the-art
static analyzer for OpenMP race detection, LLOV v.0.3
\cite{bora2021llov}, to the same suite.\footnote{While there are a
  number of effective dynamic race detectors, the goal of those tools
  is to detect races on a particular execution.  Our goal is more
  aligned with that of static analyzers: to cover as many executions
  as possible, including for different inputs, number of threads, and
  thread interleavings.}

LLOV v.0.3 implements two static analyses.  The first uses polyhedral
analysis to identify data races due to loop-carried dependencies
within OpenMP parallel loops \cite{bora2020llov}. It is unable to
identify data races involving critical sections, atomic operations,
master or single directives, or barriers. The second is a phase
interval analysis to identify statements or basic blocks (and
consequently memory accesses within those blocks) that may happen in
parallel \cite{bora2021llov}. Phases are separated by explicit or
implicit barriers and the minimum and maximum phase in which a
statement or basic block may execute define the phase interval. The
phase interval analysis errs in favor of reporting accesses as potentially
happening in parallel whenever it cannot prove that they do not; consequently, 
it may produce false alarms.

The DRB suite exercises a wide array of OpenMP language features.  Of
the 172 benchmarks, 88 use only the language primitives supported by
our CIVL OpenMP transformer (see Section \ref{sec:openmp}). Some of
the main reasons benchmarks were excluded include: use of C++,
\code{simd} and \code{task} directives, and directives for GPU
programming.  All 88 programs also use only features supported by
LLOV.  Of the 88, 47 have data races and 41 are labeled race-free.

We executed CIVL on the 88 programs, with the default number of OpenMP
threads for a parallel region bounded by $8$ (with a few exceptions,
described below).  We chose cyclic distribution as the default for
OpenMP \emph{for} loops. Many of the programs consume positive integer
inputs or have clear hard-coded integer parameters.  We manually
instrumented 68 of the 88, inserting a few lines of CIVL-C code,
protected by a preprocessor macro that is defined only when the
program is verified by CIVL.  This code allows each parameter to be
specified on the CIVL command line, either as a single value or by
specifying a range.  In a few cases (e.g., DRB055), ``magic numbers''
such as $500$ appear in multiple places, which we replaced with an
input parameter controlled by CIVL.  These modifications are
consistent with the ``small scope'' approach to verification, which
requires some manual effort to properly parameterize the program so
that the ``scope'' can be controlled.

We used the range $1..10$ for inputs, again with a few exceptions.  In
three cases, verification did not complete within 3 minutes and we
lowered these bounds as follows: for DRB043, thread bound $8$ and
input bound $4$; for the Jacobi iteration kernel DRB058, thread bound
4 and bound of $5$ on both the matrix size and number of iterations;
for DRB062, thread bound 4 and input bound $5$.

\begin{figure}[t]
\centering
\noindent
\begin{minipage}[t]{0.35\linewidth}%
\begin{lstlisting}[language=civlsmall,numbers=none,frame=none,xleftmargin=0pt]
// DRB140 (race)
int a, i;
#pragma omp parallel private(i)
{
 #pragma omp master
 a = 0;
 #pragma omp for reduction(+:a)
 for (i=0; i<10; i++)
  a = a + i;
}
\end{lstlisting}%
\end{minipage}%
\begin{minipage}[t]{0.31\linewidth}%
\begin{lstlisting}[language=civlsmall,numbers=none,frame=l,xleftmargin=2pt,escapechar=\%]
// DRB014 (race)
int n=100, m=100;
double b[n][m];
#pragma omp parallel for \
   private(j)
for (i=1;i<n;i++)
 for (j=0;j<m;j++)
  // out of bound access
  b[i][j]=b[i][j-1];
%
\end{lstlisting}%
\end{minipage}%
\begin{minipage}[t]{0.34\linewidth}%
\begin{lstlisting}[language=civlsmall,numbers=none,frame=l,xleftmargin=2pt]
// diffusion1 (race)
double *u, *v;
// alloc + init u, v
for (t=0; t<steps; t++) {
 #pragma omp parallel for
 for (i=1; i<n-1; i++) {
  u[i]=v[i]+c*(v[i-1]+v[i]);
 }
 u=v; v=u; // incorrect swap
}
\end{lstlisting}%
\end{minipage}
\caption{Excerpts from three benchmarks with data races: two from
  DataRaceBench (left and middle) and erroneous 1d-diffusion (right).}
    \label{fig:excerpts}
\end{figure}

\begin{figure}[t]
    \centering
\noindent
\begin{minipage}{0.45\linewidth}%
\begin{lstlisting}[language=civlsmall,numbers=none,frame=none,xleftmargin=0pt]
// atomic3 (no race)
int x=0, s=0;
#pragma omp parallel sections \
   shared(x,s) num_threads(2)
{
 #pragma omp section
 {
  x=1;
  #pragma omp atomic write seq_cst
  s=1;
 }
 #pragma omp section
 {
  int done = 0;
  while (!done) {
   #pragma omp atomic read seq_cst
   done = s;
  }
  x=2;
 }
}
\end{lstlisting}%
\end{minipage}%
\begin{minipage}{0.55\linewidth}
\begin{lstlisting}[language=civlsmall,numbers=none,frame=l,xleftmargin=5pt]
// bar2 (no race)
// ...create/initialize locks l0, l1;
#pragma omp parallel num_threads(2)
{
  int tid = omp_get_thread_num();
  if (tid == 0) omp_set_lock(&l0);
  else if (tid == 1) omp_set_lock(&l1);
  #pragma omp barrier
  if (tid == 0) x=0;
  if (tid == 0) {
    omp_unset_lock(&l0);
    omp_set_lock(&l1);
  } else if (tid == 1) {
    omp_set_lock(&l0);
    omp_unset_lock(&l1);
  }
  if (tid == 1) x=1;
  #pragma omp barrier
  if (tid == 0) omp_unset_lock(&l1);
  else if (tid == 1) omp_unset_lock(&l0);
}
\end{lstlisting}
\end{minipage}
    \caption{Code for synchronization using an atomic variable (left) and a 2-thread
      barrier using locks (right).}
    \label{fig:snippets}
\end{figure}

%% \newcommand{\cw}{4em}
%% \newcommand{\cs}{\hspace{1em}}
%% \newcommand{\css}{\hspace{2em}}
%% \begin{figure}[t]
  %% \centering
  %% \begin{tabular}{c@{\css}c@{\cs}c@{\cs}c@{\cs}c@{\cs}c@{\cs}|@{\cs}c@{\cs}c@{\cs}c@{\cs}c@{\cs}c}
    %% \multicolumn{1}{c@{\css}}{} & \multicolumn{5}{c@{\cs}}{DataRaceBench (88)}
    %% & \multicolumn{5}{c@{\cs}}{New Challenges (12)}\\
    %% \multicolumn{1}{c@{\css}}{} & TP & FN & TN & FP & F & TP & FN &TN & FP & F\\ \hline
    %% CIVL & 45 & 2 & 40 & 0 & 1   &   6 & 0 & 6 & 0 & 0\\
    %% LLOV & 46 & 1 & 38 & 3 & 0   &   4 & 1 & 1 & 4 & 2
  %% \end{tabular}
  %% \caption{Results of CIVL and LLOV on 100 data race benchmarks:
    %% numbers of True Positives, False Negatives, True Negatives, False
    %% Positives, and Failures.}
  %% \label{fig:results}
%% \end{figure}

CIVL correctly identified 40 of the 41 data-race-free programs,
failing only on DRB139 due to nested parallel regions.  It correctly
reported a data race for 45 of the 47 programs with data races,
missing only DRB014 (Figure~\ref{fig:excerpts}, middle) and DRB015.
In both cases, CIVL reports a bound issue for an access to
\code{b[i][j-1]} when $\code{i}>0$ and $\code{j}=0$, but fails to
report a data race, even when bound checking is disabled.

LLOV correctly identified 46 of the 47 programs with data races,
failing to report a data race for DRB140 (Figure~\ref{fig:excerpts},
left).  The semantics for reduction specify that the loop behaves as
if each thread creates a private copy, initially $0$, of the shared
variable \code{a}, and updates this private copy in the loop body.  At
the end of the loop, the thread adds its local copy onto the original
shared variable.  These final additions are guaranteed to not race
with each other.  In CIVL, this is modeled using a lock.  However,
there is no guarantee that these updates do not race with other code.
In this example, thread 0 could be executing the assignment \code{a=0}
while another thread is adding its local result to \code{a}---a data
race. This race issue can be resolved by isolating the reduction loop
with barriers.

LLOV correctly identified 38 out of 41 data-race-free programs.
It reported false alarms for DRB052 (no support for indirect
addressing), DRB054 (failure to propagate array dimensions and loop
bounds from a variable assignment), and DRB069 (failure to properly
model OpenMP lock behavior).

The DRB suite contains few examples with interesting interleaving
dependencies or pointer alias issues.  To complement the suite, we
wrote 10 additional C/OpenMP programs based on widely-used concurrency
patterns (cf.\ \cite{andrews:2000:mpd}):
\begin{itemize}
\item 3 implementations of a synchronization signal sent from one
  thread to another, using locks or busy-wait loops with critical
  sections or atomics;
\item 3 implementations of a 2-thread barrier, using busy-wait loops
  or locks;
\item 2 implementations of a 1d-diffusion simulation, one in which two
  copies of the main array are created by two separate malloc calls;
  one in which they are inside a single malloced object; and
\item an instance of a single-producer, single-consumer pattern;
  and a multiple-producer, multiple-consumer version,
  %(run with 4 producers, 4 consumers, and a buffer capacity of 5)
  both using critical sections.
\end{itemize}
For each program, we created an erroneous version with a data race,
for a total of 20 tests.  These codes are included in the experimental
archive, and two are excerpted in Figure \ref{fig:snippets}.

CIVL obtains the expected result in all 20.  While we wrote these
additional examples to verify that CIVL can reason correctly about
programs with complex interleaving semantics or alias issues, for
completeness we also evaluated them with LLOV.  It should be noted,
however, that the authors of LLOV warn that it ``\ldots{}does not
provide support for the OpenMP constructs for synchronization\ldots''
and ``\ldots{}can produce False Positives for programs with explicit
synchronizations with barriers and locks.''  \cite{bora2020llov} It is
therefore unsurprising that the results were somewhat mixed: LLOV
produced no output for 6 of our examples (the racy and race-free
versions of diffusion2 and the two producer-consumer codes) and
produced the correct answer on 7 of the remaning 14.  On these
problems, LLOV reported a race for both the racy and race-free
version, with the exception of diffusion1 (Figure~\ref{fig:excerpts},
right), where a failure to detect the alias between \texttt{u} and
\texttt{v} leads it to report both versions as race-free.

CIVL's verification time is significantly longer than LLOV's.  On the
DRB benchmarks, total CIVL time for the 88 tests was 15 minutes, 48
seconds.  Individual times ranged from $1$ to $156$ seconds: 64 took
less than 5s, 81 took less than 30s, and 82 took less than 1 minute.
(All CIVL runs used an M1 MacBook Pro with 16GB memory.)  Total CIVL
runtime on the 20 extra tests was 1 minute, 28 seconds.  LLOV analyzes
all 88 DRB problems in less than 15 seconds (on a standard Linux
machine).

%% file: sec5-related.tex
By Theorem \ref{thm:detect}, if barriers are the only form of
synchronization used in a program, only a single interleaving will be
explored, and this suffices to verify race-freedom or to find all
states at the end of each barrier epoch. This is well known in other
contexts, such as GPU kernel verification (cf.\
\cite{betts-etal:2015:gpuverify}).

Prior work involving model checking and data races for unstructured
concurrency includes Schemmel et al.\
\cite{schemmel-etal:2020:symb-por}. This work describes a technique,
using symbolic execution and POR, to detect defects in Pthreads
programs. The approach involves intricate algorithms for enumerating
configurations of prime event structures, each representing a set of
executions. The completeness results deal with the detection of
defects under the assumption that the program is race-free.  While the
implementation does check for data races, it is not clear that the
theoretical results guarantee a race will be found if one exists.

Earlier work of Elmas et al.\ describes a sound and precise technique
for verifying race-freedom in finite-state lock-based programs
\cite{elmas-qadeer-tasiran:2006:race-lockset}. It uses a bespoke
POR-based model checking algorithm that associates significant and
complex information with the state, including, for each shared memory
location, a set of locks a thread should hold when accessing that
location, and a reference to the node in the depth first search stack
from which the last access to that location was performed.

Both of these model checking approaches are considerably more complex
than the approach of this paper. We have defined a simple
state-transition system and shown that a program has a data race if
and only if a state or edge satisfying a certain condition is
reachable in that system. Our approach is agnostic to the choice of
algorithm used to check reachability. The earlier approaches are also
path-precise for race detection, i.e., for each execution path, a race
is detected if and only if one exists on that path. As we saw in the
example following Theorem \ref{thm:detect}, our approach is not
path-precise, nor does it have to be: to verify race-freedom, it is
only necessary to find one race in one execution, if one exists. This
partly explains the relative simplicity of our approach.

A common approach for verifying race-freedom is to establish
\emph{consistent correlation}: for each shared memory location, there
is some lock that is held whenever that location is accessed.
\textsc{Locksmith} \cite{pratikakis-etal:2011:locksmith} is a static
analysis tool for multithreaded C programs that takes this approach.
The approach should never report that a racy program is race-free, but
can generate false alarms, since there are race-free programs that are
not consistently correlated.  False alarms can also arise from
imprecise approximations of the set of shared variables, alias
analysis, and so on.  Nevertheless, the technique appears very
effective in practice.
% ~\cite{pratikakis:2006:locksmith}

Static analysis-based race-detection tools for OpenMP include
OMPRacer \cite{swain2020ompracer}.  OMPRacer constructs a static graph
representation of the happens-before relation of a program and
analyzes this graph, together with a novel whole-program pointer
analysis and a lockset analysis, to detect races.  It may miss
violations as a consequence of unsound decisions that aim to improve
performance on real applications. The tool is not open source.  The
authors subsequently released OpenRace~\cite{swain2021openrace},
designed to be extensible to other parallelism dialects; similar to
OMPRacer, OpenRace may miss violations.  Prior papers by the
authors present details of static methods for race detection, without
a tool that implements these methods~\cite{swain2018towards}.

PolyOMP~\cite{chatarasiPolyOMP2017} is a static tool that uses a
polyhedral model adapted for a subset of OpenMP. Like most polyhedral
approaches, it works best for affine loops and is precise in such
cases. The tool additionally supports may-write access relations for
non-affine loops, but may report false alarms in that
case. DRACO~\cite{ye2018using} also uses a polyhedral model and has
similar drawbacks.

Hybrid static and dynamic tools include
Dynamatic~\cite{davis2021dynamatic}, which is based on LLVM.  It
combines a static tool that finds candidate races, which are
subsequently confirmed with a dynamic tool.  Dynamatic may report
false alarms and miss violations.

ARCHER~\cite{atzeni-gopalakrishnan-etal:2016:archer} is a tool that
statically determines many sequential or provably non-racy code
sections and excludes them from dynamic analysis, then uses
TSan~\cite{serebryany-iskhodzhanov:2009:tsan} for dynamic race
detection.  To avoid false alarms, ARCHER also encodes information
about OpenMP barriers that are otherwise not understood by TSan. A
follow-up paper discusses the use of the OMPT interface to aid dynamic
race detection tools in correctly identifying issues in OpenMP
programs~\cite{protze-hahnfeld-etal:2017:ompt}, as well as
SWORD~\cite{atzeni-gopalakrishnan-etal:2018:sword}, a dynamic tool
that can stay within user-defined memory bounds when tracking races,
by capturing a summary on disk for later analysis.

ROMP~\cite{gu-mellor-crummey:2018:romp} is a dynamic/static tool that
instruments executables using the DynInst library to add checks for
each memory access and uses the OMPT interface at runtime.  It claims
to support all of OpenMP except \code{target} and \code{simd}
constructs, and models ``logical'' races even if they are not
triggered because the conflicting accesses happen to be scheduled on
the same thread. Other approaches for dynamic race detection and
tricks for memory and run-time efficient race bookkeeping during
execution are described in~\cite{mellor:1991:on-the-fly,
  ha-jun-etal:2011:enr-labeling, ha-kuh-etal:2012:on-the-fly,
  boushehrinejadmoradi-yoga:2020:on-the-fly}.

% ARBALEST~\cite{lechen-protze-etal:2021:arbalest} is a tool to detect
% data mapping issues that occur when incorrectly transferring (or
% failing to transfer) data between a host and target device. Other
% work presents tools to analyze the impact of data
% races~\cite{wang-lin:2021:does-it-matter}, or focuses on OpenMP
% tasks~\cite{matar-unat:2018:tasks}.

%% Test problems used:
%% \begin{itemize}
%% \item PolyOMP: 34 OpenMP programs from the OmpSCR and PolyBench-ACC
%% \item DRACO: DataRaceBench and AMG2013
%% \item LLOV-MHP: DatRaceBench and 22 applications (many from ECP proxy apps)
%% \item ROMP (hybrid): DataRaceBench and OmpSCR
%% \end{itemize}

Deductive verification approaches have also been applied to OpenMP
programs.  An example is \cite{blom2021correct}, which introduces an
intermediate parallel language and a specification language based on
permission-based separation logic.  C programs that use a subset of
OpenMP are manually annotated with ``iteration contracts'' and then
automatically translated into the intermediate form and verified using
VerCors and Viper.  Successfully verified programs are guaranteed to
be race-free.  While these approaches require more work from the user,
they do not require bounding the number of threads or other
parameters.

%% file: sec6-conclude.tex
In this paper, we introduced a simple model-checking technique to
verify that a program is free from data races.  The essential ideas
are (1) each thread ``remembers'' the accesses it performed since its
last synchronization operation, (2) a partial order reduction scheme
is used that treats all memory accesses as local, and (3) checks for
conflicting accesses are performed around synchronizations.  We proved
our technique is sound and precise for finite-state models, using a
simple mathematical model for multithreaded programs with locks and
barriers.  We implemented our technique in a prototype tool based on
the CIVL symbolic execution and model checking platform and applied it
to a suite of C/OpenMP programs from DataRaceBench. Although based on
completely different techniques, our tool achieved performance
comparable to that of the state-of-the-art static analysis tool, LLOV
v.0.3.

Limitations of our tool include incomplete coverage of the OpenMP
specification (e.g., \code{target}, \code{simd}, and \code{task}
directives are not supported); the need for some manual
instrumentation; the potential for state explosion necessitating small
scopes; and a combinatorial explosion in the mappings of threads to
loop iterations, OpenMP sections, or single constructs.  In the last
case, we have compromised soundness by selecting one mapping, but in
future work we will explore ways to efficiently cover this space.
On the other hand, in contrast to LLOV and because of the reliance on model
checking and symbolic execution, we were able to verify the presence
or absence of data races even for programs using unstructured
synchronization with locks, critical sections, and atomics, including
barrier algorithms and producer-consumer code.

%% file: appendix.tex
% Steve
% Outline
% A. A Reduction Thm.  Strategy, assumptions on barriers.
% A.1.  Statement of the Theorem
% A.2.  Key Lemmas
% A.3.  Proof of the Theorem
% 
% B. Proof of Detection Thm.
% B.1. Preliminatires
% B.2. Block Decomposition
% B.3. Blocks and Data Races
% B.4. Proof: first in case of no barriers.  then in case of barriers.

\section{A Reduction Theorem}

The goal of this appendix is to prove Theorem \ref{thm:detect}.  To do
this, we will first consider programs without barriers, as this
simplifies many aspects of the argument.  Then we show that Theorem
\ref{thm:detect} can be easily obtained from the result about programs
without barriers.

In this section, we prove a general reduction theorem for programs
without barriers.  This theorem follows the standard ``ample set''
approach to partial order reduction.  It is shown that if subsets of
enabled transitions satisfy specific axioms, then the reduction is
sound for detecting data races.  The reduction theorem will be a key
step in the proof of Theorem \ref{thm:detect}.

\subsection{Statement of the Reduction Theorem}

Let $P$ be a program without barriers.  This means
$\BarrierState_i=\emptyset$ for all $i\in\TID$.  No $\barrierExit_i$
statement occurs in any execution.  The wait set component of the
state ($2^{\TID}$) has no impact on the $\enabled$ or $\execute$
functions and can be ignored.

\begin{definition}
  \label{def:space}
  A \emph{\stategraph} of $P$ is a triple $\graph=(V,E,\state)$, where
  \begin{enumerate}
  \item $V$ is a set of \emph{nodes},
  \item $E\subseteq V\times \Stmt\times V$ is a set of \emph{edges},
  \item $\state\colon V\ra\State$,
  \item if $(u,t,v)\in E$ then $\state(u)\stackrel{t}{\ra}\state(v)$
    is a transition,
  \item for all $u,v,v'\in V$ and $t\in\Stmt$,
    $(u,t,v),(u,t,v')\in E \implies v=v'$.   \qed
  \end{enumerate}
\end{definition}

Fix a state graph $\graph=(V,E,\state)$ of $P$.

For $u,v\in V$ and $t\in\Stmt$, write $u\stackrel{t}{\ra}v$ for
$(u,t,v)$, and define
\begin{align*}
  \enabled(u) &= \enabled(\state(u))\\
  \ample_\graph(u) &= \{t\in\Stmt\mid \exists v\in V \;.\;
                u\stackrel{t}{\ra}v\in E\}.
\end{align*}
Clearly, $\ample_\graph(u)\subseteq\enabled(u)$.  Also, if
$t\in\ample_\graph(u)$, then there is a unique $v\in V$ such that
$u\stackrel{t}{\ra}v\in E$.

\begin{definition}
  \label{def:path}
  A \emph{path in $\graph$} is a finite or infinite sequence of nodes
  and edges
  \( \alpha=(u_0\stackrel{t_1}{\ra}u_1\stackrel{t_2}{\ra}\cdots) \)
  such that $u_0\stackrel{t_1}{\ra}u_1\in E$,
  $u_1\stackrel{t_2}{\ra}u_2\in E$, \ldots.  The \emph{execution
    defined by $\alpha$} is the execution
  \( \overline{\alpha} = (\state(u_0)\stackrel{t_1}{\ra}
  \state(u_1)\stackrel{t_2}{\ra}\cdots).  \) \qed
\end{definition}
The fact that $\overline{\alpha}$ is an execution follows
from Definition \ref{def:space}(4).

\begin{definition}
  \label{def:dense}
  We say $\graph$ is \emph{dense} if all of the following hold:
  \begin{enumerate}
  \item for all $u\in V$, if $\enabled(u)\neq\emptyset$ then
    $\ample_\graph(u)\neq\emptyset$
  \item for all $u\in V$, if $t\in\ample_\graph(u)$, $t'\in\enabled(u)$,
    and $\tid(t)=\tid(t')$ then $t'\in\ample_\graph(u)$
  \item for all $u\in V$, if $\ample_\graph(u)\neq\enabled(u)$ then
    all statements in $\ample_\graph(u)$ are nsync statements
  \item for any cycle in $\graph$ (a path of positive length from some node
    to itself), there is some $v\in V$ on the cycle with
    $\ample_\graph(v)=\enabled(v)$.   \qed
  \end{enumerate}
\end{definition}

We can now state the Reduction Theorem:

\begin{theorem}
  \label{thm:reduce}
  Let $P$ be a multithreaded program, $\graph=(V,E,\state)$ a dense
  \stategraph{} for $P$, and $u_0\in V$.  Assume the set of nodes
  reachable from $u_0$ is finite.
  \begin{enumerate}
  \item  If there is an execution from
    $\state(u_0)$ with a data race then there is a path $\beta$ in
    $\graph$ from $u_0$ such that $\overline{\beta}$ has a data race.
  \item If there is an execution from $\state(u_0)$ to a state $s_f$
    with $\enabled(s_f)=\emptyset$, then there is a path in $\graph$
    from $u_0$ to a node $u_f$ with $\state(u_f)=s_f$.
  \end{enumerate}
\end{theorem}

\subsection{Key Lemmas}

In this section, we prove some basic lemmas involving commuting
transitions and the data race relation.  These will be used in the
proof of Theorem \ref{thm:reduce}.  We continue assuming $P$ has no
barriers.

Suppose $t$ is an nsync statement.  It follows from the definitions
that no statement from another thread can enable or disable $t$.
Moreover, $t$ cannot enable or disable any statement from another
thread.  This is made precise as follows:
\begin{lemma}
  \label{lem:nsync-enable}
  Let $t_1$ be an nsync statement, $t_2\in\Stmt$, and $s\in\State$.
  Assume $\tid(t_1)\neq\tid(t_2)$.  Then
  \begin{enumerate}
  \item if $t_2\in\enabled(s)$ then $t_1\in\enabled(s)\iff
    t_1\in\enabled(\execute(s, t_2))$
  \item if $t_1\in\enabled(s)$ then $t_2\in\enabled(s)\iff
    t_2\in\enabled(\execute(s, t_1))$.
  \end{enumerate}
\end{lemma}
\begin{proof}
  Let $p=\tid(t_1)$ and $q=\tid(t_2)$.
  Say $s=\langle \xi,\zeta,\theta\rangle$.

  Proof of (1): Assume $t_2\in\enabled(s)$.  Let
  $s'=\execute(s, t_2)$.  Say $s'=\langle\xi',\zeta',\theta'\rangle$.
  We have $\xi'_p=\xi_p$, i.e., executing a statement in thread $q$
  does not change the local state of thread $p$.  As
  $\xi_p\in\NSyncState$,
  \[
    \enabled_p(\xi_p,\zeta,\theta)=\enabled_p(\xi_p,\zeta',\theta').
  \]
  Hence
  \[
    \Stmt_p\cap\enabled(s)=\enabled_p(\xi_p,\zeta,\theta)
    =\enabled_p(\xi_p',\zeta',\theta')
    =\Stmt_p\cap\enabled(s').
  \]
  It follows that $t_1\in\enabled(s)\iff t_1\in\enabled(s')$.

  Proof of (2): Assume $t_1\in\enabled(s)$.  Let
  $s'=\execute(s, t_1)$.  As $t_1$ is an nsync statement, it does not
  change the state of the locks.  Say
  $s'=\langle\xi',\zeta',\theta\rangle$.  We have $\xi'_q=\xi_q$ since
  executing a statement in thread $p$ does not change the local state
  of thread $q$.  If $t_2$ is an acquire or release statement, then
  $t_2\in\enabled(s)\iff t_2\in\enabled(s')$ since these depend only
  on the local state of $q$ and the state of the locks.  If $t_2$ is
  an nsync statement, then the desired result follows from part (1),
  swapping $t_1$ and $t_2$. \qed
\end{proof}

\begin{lemma}
  \label{lem:extend-race}
  Let $\alpha'$ be an execution and $\alpha$ a prefix of $\alpha'$.
  Then $\DR(\alpha)\subseteq\DR(\alpha')$.
\end{lemma}
\begin{proof}
  Note $\tr(\alpha)$ is a prefix of $\tr(\alpha')$ and
  $[\alpha]\subseteq [\alpha']$.  Suppose $e,f\in[\alpha]$ and
  $(e,f)\in\HB(\alpha')$.  We will show $(e,f)\in\HB(\alpha)$.

  Since $\HB(\alpha')$ is a transitive closure, there is a finite
  sequence of elements of $[\alpha']$
  \[
    e=e_1, e_2, \ldots, e_r=f
  \]
  such that for each $i$ ($1\leq i<r$) either $(e_i,e_{i+1})$ is in
  the intra-thread order relation or in the release-acquire relation.
  In either case, $e_i$ occurs before $e_{i+1}$ in the sequence
  $\tr(\alpha')$.  Since $f\in[\alpha]$, that means all
  $e_i\in[\alpha]$.  It follows that $(e,f)\in\HB(\alpha)$.

  Suppose $e$ and $e'$ race in $\alpha$.  Then neither happens before
  the other in $\alpha$.  By the paragraph above, neither happens
  before the other in $\alpha'$.  Since the statements of $e$ and
  $e'$ conflict, $e$ and $e'$ race in $\alpha'$. \qed
\end{proof}

\begin{lemma}
  \label{lem:commute}
  Let
  $\alpha = (s_0\stackrel{t_1}{\ra}\cdots\stackrel{t_{n}}{\ra}s_n)$
  be a finite execution with trace $e_1\cdots e_{n}$.  Suppose
  $1\leq i\leq n-1$, $(e_i,e_{i+1})\not\in\HB(\alpha)$, and
  $(t_i,t_{i+1})\not\in\conflict$.  Then there exists $s'\in\State$
  such that
  \[
    \alpha'=(s_0\stackrel{t_1}{\ra}\cdots
    \stackrel{t_{i-1}}{\ra}s_{i-1}
    \stackrel{t_{i+1}}{\ra}s'
    \stackrel{t_i}{\ra}s_{i+1}
    \stackrel{t_{i+2}}{\ra}\cdots
    \stackrel{t_{n}}{\ra}s_n)
  \]
  is an execution.  Moreover, $[\alpha]=[\alpha']$,
  $\HB(\alpha)=\HB(\alpha')$, and $\DR(\alpha)=\DR(\alpha')$.
\end{lemma}

\begin{proof}
  We have $t_{i+1}\in\enabled(\execute(s_{i-1},t_i))$.  Let
  $p=\tid(t_i)$ and $q=\tid(t_{i+1})$.  Since
  $(e_i,e_{i+1})\not\in\HB(\alpha)$, $p\neq q$.  We claim all of the
  following hold:
  \begin{align}
    \label{eq:claim1} t_{i+1}&\in\enabled(s_{i-1})\\
    \label{eq:claim2} t_i&\in\enabled(\execute(s_{i-1},t_{i+1}))\\
    \label{eq:claim3}
    \execute(\execute(s_{i-1},t_{i+1}), t_i) &=
    \execute(\execute(s_{i-1},t_i),t_{i+1}).
  \end{align}
  If the claim holds, take $s'=\execute(s_{i-1}, t_{i+1})$, and the
  existence of $\alpha$ follows.  The proof of the claim is in two
  cases.
  
  \textbf{Case 1:} $t_{i+1}$ is an nsync statement.  Then
  \eqref{eq:claim1} follows from Lemma \ref{lem:nsync-enable}(1), and
  \eqref{eq:claim2} follows from Lemma \ref{lem:nsync-enable}(2).  If
  $t_i$ is a lock statement, then \eqref{eq:claim3} follows from the
  definition of $\execute$, as the two statements modify distinct
  components of the state.  If $t_i$ is an nsync statement, then
  \eqref{eq:claim3} follows from the assumption that
  $(e_i,e_{i+1})\not\in\conflict$.

  \textbf{Case 2}: $t_{i+1}$ is a lock statement.  If $t_i$ is an
  nsync statement then the claim follows by an argument similar to
  that of Case 1.  So suppose $t_i$ is also a lock statement.  We
  claim that the two lock statements operate on different locks.  If
  both statements are acquires, then they must operate on different
  locks, as an acquire statement is only enabled when the lock is
  free.  If $t_i$ is an acquire and $t_{i+1}$ a release, then again
  they must operate on different locks, else $p=q$, as only the thread
  owning the lock can perform a release operation on that lock.  If
  $t_i$ is a release and $t_{i+1}$ an acquire then again they must
  operate on different locks, else $(e_i,e_{i+1})\in\HB(\alpha)$.
  Finally, if both statements are releases, then they must operate on
  different locks, since a release statement is only enabled when the
  lock is owned by some thread.  The claim is now clear, since the two
  statements are operating on distinct components of the lock state.

  We have $[\alpha]=[\alpha']$ since $p\neq q$.  We must show
  $\HB(\alpha)=\HB(\alpha')$.  The intra-thread relation is the same
  in $\alpha$ and $\alpha'$.  The release-acquire relation is also the
  same since it is not the case that one statement is an acquire
  statement and the other a release statement on the same lock.  Hence
  the transitive closure of the union of the two relations is
  identical.  Finally, the data race relation depends only on
  happens-before and \conflict{}, so $\DR(\alpha)=\DR(\alpha')$. \qed
\end{proof}

\begin{lemma}
  \label{lem:ncommute}
  Let
  $\alpha = (s_0\stackrel{t_1}{\ra}\cdots\stackrel{t_{n}}{\ra}s_n)$ be
  a finite execution.  Suppose $1\leq i\leq n-1$, and $t_i$ and
  $t_{i+1}$ are conflicting nsync statements from different threads.
  Then there exist $s',s''\in\State$ such that
  \[
    \alpha'=(s_0\stackrel{t_1}{\ra}\cdots
    \stackrel{t_{i-1}}{\ra}s_{i-1}
    \stackrel{t_{i+1}}{\ra}s'
    \stackrel{t_i}{\ra}s'')
  \]
  is an execution.  In addition, $\alpha'$ contains a data race.
\end{lemma}

\begin{proof}
  Say $\tr(\alpha)=e_1\cdots e_n$. The existence of $s'$ and $s''$
  follows from Lemma \ref{lem:nsync-enable}.  Since $t_i$ and
  $t_{i+1}$ are in different threads,
  $\tr(\alpha')=e_1\cdots e_{i-1}e_{i+1}e_i$.  Clearly neither
  $(e_i,e_{i+1})$ nor $(e_{i+1},e_i)$ is in $\HB(\alpha')$.  As
  $(t_i,t_{i+1})\in\conflict$, $(e_i,e_{i+1})\in\DR(\alpha')$.
  \qed
\end{proof}

\subsection{Proof of the Reduction Theorem}

We now complete the proof of Theorem \ref{thm:reduce}.  Let
$s_0=\state(u_0)$ and
\[
  \alpha=(s_0\stackrel{t_1}{\ra}\cdots\stackrel{t_{n}}{\ra}s_n)
\]
be a finite execution.

We first prove part 1.  Suppose $\alpha$ contains a data race.  We
construct a path in $\graph$ with a data race.  To do this, we show
there exists a sequence of pairs $(\beta_i,\gamma_i)_{i\geq 0}$
satisfying all of the following for all $i\geq 0$:
\begin{enumerate}
\item $\beta_i$ is a path in $\graph$ starting at $u_0$
\item $|\beta_i|=i$
\item $\beta_i$ is a prefix of $\beta_{i+1}$
\item $\gamma_i$ is an execution
\item the final state of $\overline{\beta}_i$ is the initial state of
  $\gamma_i$
\item $\overline{\beta}_i\circ\gamma_i$ has a data race
\item $|\gamma_{i+1}|\leq|\gamma_i|$.
\end{enumerate}
In addition we will show there exists $m\geq 0$ such that
\[
  |\gamma_{m}|=0.
\]
It follows that $\beta_m$ is a path in $\graph$ starting at $u_0$ and
$\overline{\beta}_m$ has a data race.

Let $\beta_0$ be the path of length $0$ at $u_0$, and
$\gamma_0=\alpha$.

Suppose $i\geq 0$ and $\beta_i$ and $\gamma_i$ have been constructed
to satisfy (1)--(7).  If $|\gamma_i|=0$, let $\beta_{i+1}=\beta_i$ and
$\gamma_{i+1}=\gamma_i$; clearly (1)--(7) still hold.

So assume $|\gamma_i|>0$.  We construct $\beta_{i+1}$ and
$\gamma_{i+1}$ as follows.

Let $u$ be the final node of $\beta_i$, so $s=\state(u)$ is the
initial state of $\gamma_i$.  Since $|\gamma_i|>0$, there is at least
one enabled statement at $s$.  By Definition \ref{def:dense}(1),
$\ample_\graph(u)\neq\emptyset$.

\subsubsection{Case 0: $\ample_\graph(u)=\enabled(u)$.}
Say $\gamma_i = (s\stackrel{t}{\ra}s')\circ\gamma'$.  Since
\[
  t\in\enabled(s)=\enabled(u)=\ample_\graph(u),
\]
there is a unique $v\in V$ such that $u\stackrel{t}{\ra}v$.  Let
$\beta_{i+1}=\beta_i\circ(u\stackrel{t}{\ra}v)$ and
$\gamma_{i+1}=\gamma'$.  Hence
\[
  \overline{\beta}_{i+1}\circ\gamma_{i+1} =
  \overline{\beta}_i\circ(s\stackrel{t}{\ra}s')\circ\gamma' =
  \overline{\beta}_i\circ\gamma_i.
\]
It is clear that conditions (1)--(7) hold.  Moreover
$|\gamma_{i+1}|<|\gamma_i|$.  We say that $(\beta_{i+1},\gamma_{i+1})$
are formed by performing a \emph{shift} on $(\beta_i,\gamma_i$).

\subsubsection{Case 1: $\ample_\graph(u)\neq\enabled(u)$.}
In this case, $\ample_\graph(u)$ consists of nsync statements.  We
explore two sub-cases.

\vspace{1ex}\par\noindent\textbf{Case 1a: $\gamma_i$ contains a
  statement in $\ample_\graph(u)$.}  Let $t$ be the first statement in
$\ample_\graph(u)$ in $\gamma_i$.  Let $p=\tid(t)$.  We claim $t$ is the
first statement in $\gamma_i$ from thread $p$.  To see this, let $t'$
be the first statement in $\gamma_i$ from $p$. Since $p$ is at an
nsync state in $s$, $t'$ is an nsync statement.  By Lemma
\ref{lem:nsync-enable}, $t'$ is enabled at $s$.  By Definition
\ref{def:dense}(2), $t'\in\ample_\graph(u)$.  Hence $t'=t$.

We now transform $\gamma_i$ by repeatedly transposing $t$ with the
statement to its left.  Let $k$ be the index of $t$ in $\gamma_i$. We
construct a sequence of executions $\gamma_{i,j}$ ($0\leq j\leq k$).
For each $j$, conditions (4)--(6) will hold with $\gamma_{i,j}$ in
place of $\gamma_i$, and (7) will hold with $\gamma_{i,j}$ in place of
$\gamma_{i+1}$. In addition, $t$ will occur in index $j$ of
$\gamma_{i,j}$.

We start at $k$ and work down to $0$.  Let $\gamma_{i,k}=\gamma_i$.
Suppose $j\geq 1$ and $\gamma_{i,j}$ has been defined.  We will define
$\gamma_{i,j-1}$.  Let $t'$ be the statement at position $j-1$ in
$\gamma_{i,j}$.

If $t$ does not conflict with $t'$, then by Lemma \ref{lem:commute},
$t$ and $t'$ may be transposed to yield a new execution
$\gamma_{i,j-1}$, and $\overline{\beta}_i\circ\gamma_{i,j-1}$ has the
same data race relation as $\overline{\beta}_i\circ\gamma_{i,j}$,
which has a race.

If $t$ and $t'$ are conflicting nsync statements, by Lemma
\ref{lem:ncommute} there is an execution $\gamma_{i,j-1}$ which is the
prefix of length $j+1$ of the result of transposing $t$ and $t'$, and
$\overline{\beta}_i\circ\gamma_{i,j-1}$ again has a race.

Now $\gamma_{i,0}$ has $t$ in position $0$.  Let
$(\beta_{i+1},\gamma_{i+1})$ be the result of performing a shift on
$(\beta_i,\gamma_{i,0})$.  Note $|\gamma_{i+1}|<|\gamma_i|$.

\vspace{1ex}\par\noindent\textbf{Case 1b: $\gamma_i$ does not contain
  a statement in $\ample_\graph(u)$.}
In this case we will insert a new statement, appending it to
$\beta_i$.

Choose any $t\in\ample_\graph(u)$ and let $p=\tid(t)$.  There is no
statement in $\gamma_i$ from thread $p$.  (As argued above, if there
were such a statement, then the first statement in $\gamma_i$ from
$p$ would be in $\ample_\graph(u)$.)

Since $t$ is an nsync statement, it cannot be disabled by statements
from other processes.  Hence we can extend $\gamma_i$ to
$\tilde{\gamma}_i$ by appending $t$ and one more state.  Note
$\overline{\beta}_i\circ\tilde{\gamma}_i$ also contains a data race.
Now we can apply the technique of Case 1a to $\tilde{\gamma}_i$ to
move $t$ to position $0$ while maintaining the race, and then perform
a shift.  In the worst case, $|\gamma_{i+1}|=|\gamma_{i}|$.

\subsubsection{Termination.}

We have to show that for any $i\geq 0$, there is some
$j>i$ such that the construction of $(\beta_j,\gamma_j)$
does not involve Case 1b.

So suppose there is some $i\geq 0$ such that for all $j\geq i$, the
construction of $(\beta_j,\gamma_j)$ involves Case 1b.  The paths
\[
  \beta_i,\beta_{i+1},\beta_{i+2},\ldots
\]
satisfy (1)--(3).  For $j\geq i$, let $u_j$ be the terminal node of
$\beta_j$.

Since all the $u_j$ are reachable from $u_0$, and the set of nodes
reachable from $u_0$ is finite, there exist integers $j,k$ such that
$i\leq j<k$ and $u_j=u_k$.  Hence $u_j,u_{j+1},\ldots,u_k$ form a
cycle in $\graph$.  By Definition \ref{def:dense}(4), there is some
$l$ with $j\leq l<k$ and $\ample_\graph(u_l)=\enabled(u_l)$.  But
then, the construction of $(\beta_{l+1},\gamma_{l+1})$ would use Case
0, a contradiction.

This completes the proof of part 1 of Theorem \ref{thm:reduce}.

We now turn to the proof of part 2.  Suppose $\alpha$ does not contain
a data race, but ends at a state with no enabled statement.  The proof
is mostly the same as that of part 1.  Rather than repeat the proof,
we summarize the parts that change.

First, replace invariant 6 ($\overline{\beta}_i\circ\gamma_i$ has a
data race) with the following: the final state of execution $\gamma_i$
is the final state of $\alpha$.  As $\gamma_0=\alpha$, this holds for
$i=0$.

The second change is that Case 1b of the inductive step never
occurs. Recall that in Case 1, $u$ is the final node of path
$\beta_i$, $s=\state(u)$ is the initial node of $\gamma_i$,
$\ample_G(u)\neq\emptyset$, and $\ample_G(u)\neq\enabled(u)$.  We wish
to show that $\gamma_i$ contains a statement in $\ample_G(u)$.  Let
$t$ be any element of $\ample_G(u)$ and $p=\tid(t)$.  Since thread $p$
is at an nsync state at $u$, but is at a terminal or acquire state at
the end of $\alpha$ (and therefore at the end of $\gamma_i$),
$\gamma_i$ must contain a statement from $p$.  Let $t'$ be the first
statement from $p$ in $\gamma_i$. Then the nsync statement $t'$ is
also enabled at $u$, as statements from other threads cannot enable an
nsync statement, and therefore $t'\in\ample_G(u)$.  Therefore Case 1a
must hold.

The third observation is that in Case 1a, it is never the case that
two transitions being transposed conflict, because of the assumption
that $\alpha$ is data race-free.

Hence the transformation carried out in the inductive step involves
only a sequence of transpositions of adjacent commuting transitions.
As such a transposition does not change the final state, the final
state of $\gamma_i$ is invariant.  At termination, $\gamma_i$ is
empty and $\beta$ terminates at a node with state the final state of
$\alpha$.

\section{The Race Detection Theorem}

In this section, we prove Theorem \ref{thm:detect}.  We first prove
the theorem under the assumption that $P$ has no barriers.

\subsection{Preliminaries}

Recall from the discussion before Definition \ref{def:race-space} that
we assume the program $P$ comes with sets $R_i$, for each $i\in\TID$.
For each $i$,
$\UnlockingState_i\cup\LockingState_i\subseteq
R_i\subseteq\LocalState_i$, and any cycle in the local graph of thread
$i$ has at least one node in $R_i$.

\begin{definition}
  \label{def:normal}
  For $s\in\State$ and $i\in\TID$, we say thread $i$ is \emph{normal
    at $s$} if the local state of thread $i$ in $s$ is not in $R_i$
  and thread $i$ has an enabled statement at $s$.  If $v$ is a node in
  a state graph, thread $i$ is \emph{normal at $v$} if thread $i$ is
  normal at $\state(v)$.
\end{definition}
As we are assuming $P$ has no barriers, if thread $i$ is normal at $s$
then thread $i$ must be at an nsync state at $s$.

\begin{lemma}
  \label{lem:cycle}
  Let $P$ be a multithreaded program without barriers and
  $\graph=(V,E)$ the race-detecting \stategraph{} for $P$.  Any
  infinite path in $G$ with only a finite number of nodes has a node
  at which no thread is normal.
\end{lemma}
\begin{proof}
  Let $\zeta=(u_0\stackrel{t_1}{\ra}u_1\stackrel{t_2}{\ra}\cdots)$ be
  an infinite path in $G$.  For $i\geq 0$, let $m_i$ be the number of
  threads that are normal at $u_i$.  Let $a\geq 0$ be an integer for
  which $m_a=\min \{m_k\mid k\geq 0\}$.  Suppose $m_a>0$.  We will
  arrive at a contradiction.

  Let $i$ be the minimal ID of a normal thread at $u_a$.  We show by
  induction that for all $b\geq a$, $m_b=m_a$ and thread $i$ is the
  minimal ID of a normal thread at $u_b$.  The inductive hypothesis
  clearly holds for $b=a$.
  
  Suppose $b>a$ and the inductive hypothesis holds for $b-1$.  Then
  $m_{b-1}=m_a$ and $i$ is the minimal ID of a normal thread at
  $u_{b-1}$.  Moreover, $t_{b}$ is an nsync statement in thread $i$,
  by Definition \ref{def:race-space}.  For $j\in\TID\setminus\{i\}$,
  thread $j$ is normal at $u_{b-1}$ if and only if thread $j$ is
  normal at $u_{b}$, as nsync statements cannot be enabled or
  disabled by actions from other threads.  Since $a$ was chosen to
  minimize the $m_k$, we must have $m_{b}=m_{b-1}=m_a$ and $i$ is
  still the minimal thread ID of a thread that is normal at $u_{b}$.
  This proves the inductive step.

  Projecting onto the local state of thread $i$, the suffix of $\zeta$
  starting from $u_a$ yields an infinite path in the local graph of
  thread $i$ with a finite number of states, but which never passes
  through a state in $R_i$.  This path must contain a cycle,
  contradicting the assumption that every cycle in the local graph has
  a state in $R_i$. \qed
\end{proof}

\begin{lemma}
  \label{lem:dense}
  Let $P$ be a multithreaded program without barriers and
  $\graph=(V,E)$ a race-detecting \stategraph{} for $P$.  Define
  $\state\colon V\ra\State$ by $\state(\langle s,\ma\rangle)=s$.  Then
  $(V,E,\state)$ is a dense \stategraph.
\end{lemma}

\begin{proof}
  Let $\langle s,\ma\rangle\in V$ and
  $T=\ample_\graph(\langle s,\ma\rangle)$.  There are two cases: In
  the first case, a thread is normal at $s$.  Then $T$ consists of all
  enabled statements in thread $i$, where $i$ is the minimal ID of
  such a thread.  In the second case, there is no normal thread at
  $s$.  Then $T$ consists of all enabled statements.  In either case,
  if there is an enabled statement at $\langle s,\ma\rangle$ then $T$
  is nonempty.  Hence Definition \ref{def:dense}(1) holds.

  In the first case, $T$ consists of all enabled statements in one
  thread.  In the second case $T$ consists of all enabled statements
  in all threads.  Hence Definition \ref{def:dense}(2) holds.

  If $T$ is a proper subset of the enabled statements then the first
  case holds.  In this case, the statements of $T$ come from an nsync
  state, hence are all nsync statements.  So Definition
  \ref{def:dense}(3) holds.

  If $\alpha$ is a cyclic path in $\graph$, then there is an infinite
  path in $\graph$ with a finite number of nodes, formed by repeating
  $\alpha$ infinitely.  By Lemma \ref{lem:cycle}, there is a node $u$
  on $\alpha$ at which no thread is normal.  According to Definition
  \ref{def:race-space}, $u$ is fully enabled, satisfying Definition
  \ref{def:dense}(4). \qed
\end{proof}

One direction of the proof of Theorem \ref{thm:detect} is
straightforward:

\begin{lemma}
  \label{lem:sound}
  If $\graph$ detects a race from $u_0$ then $P$ has an execution
  starting from $s_0$ with a data race.
\end{lemma}
\begin{proof}
  Let $u$ be the target node of an edge in $\graph$ that detects a
  race.

  Let $t_1$ and $t_2$ be a pair of conflicting statements stored at
  $u$, and $i=\tid(t_1)$ and $j=\tid(t_2)$.  There is a path $\zeta$
  from $u_0$ that terminates at $u$.  At least one edge in $\zeta$ is
  labeled with $t_1$; let $e_1$ be the last such edge.  When $e_1$
  executes, $t_1$ is added to $\ma_i$ and is not removed by any
  statement on $\zeta$ after that point.  Since any release statement
  in thread $i$ removes all statements from $\ma_i$, no release
  statement in $\zeta$ occurs in thread $i$ after $t_1$. Define $e_2$
  similarly.

  In the trace resulting from $\zeta$, the events corresponding to
  $e_1$ and $e_2$ cannot be ordered by happens-before, because there
  is no release event in thread $i$ after $e_i$, and no release event
  in thread $j$ after $e_j$.  Hence the path defines an execution with
  a data race. \qed
\end{proof}

The other direction is more involved.

\subsection{Block Decomposition}

We continue with our assumption that $P$ has no barriers.  Let
$\graph=(V,E,\state)$ be the race-detecting \stategraph{}.

Given a finite path
$\alpha=(u_0\stackrel{t_1}{\ra}\cdots\stackrel{t_{n}}{\ra}u_{n})$ in
$\graph$, we define integers $N$ and $i_0, \ldots, i_{N}$ as follows.
Let $i_0=0$.  Assume $j\geq 0$ and $i_j$ has been defined.  If $i_j=n$
then $N=j$.  If $i_j<n$, define $i_{j+1}$ by

\begin{align*}
  C &=
      \begin{aligned}[t]
        \{k\in [i_j+2,n] \mid\ 
        & \text{$\tid(t_k)\neq\tid(t_{i_j+1})$ or}\\
        & \text{thread $\tid(t_k)$ is not normal at $u_{k-1}$}\}
      \end{aligned}
  \\
  i_{j+1}
    &=
      \begin{cases}
        \min(C)-1 & \text{if $C\neq \emptyset$}\\
        n & \text{otherwise}.
      \end{cases}
\end{align*}

\begin{definition}
  \label{def:block}
  Let
  $\alpha=(u_0\stackrel{t_1}{\ra}\cdots\stackrel{t_{n}}{\ra}u_{n})$
  be a finite path in $\graph$.  Define $N$ and $i_0, \ldots, i_{N}$
  as above. The \emph{block length} of $\alpha$ is $N$.  For
  $1\leq j\leq N$, let
  \[
    B_j = (u_{i_{j-1}}\stackrel{t_{i_{j-1}+1}}{\ra}\cdots
    \stackrel{t_{i_{j}}}{\ra}u_{i_{j}}).
  \]
  The path $B_j$ is a \emph{block} of $\alpha$.  Define
  $\tid(B_j)=\tid(t)$, where $t=t_{i_{j-1}+1}$ is the first statement
  of $B_j$.  $B_j$ is \emph{initial} if thread $\tid(B_j)$ is normal
  at $u_{i_{j-1}}$.  \qed
\end{definition}

Note: a path of length $0$ has block length $0$.  A path of positive
length has a positive block length.

\begin{lemma}
  \label{lem:block-omnibus}
  Let $\alpha$ be a finite path in $\graph$, and $B_1,\ldots,B_{N}$
  the blocks of $\alpha$.  All of the following hold:
  \begin{enumerate}
  \item every block has length at least one and all statements
    in the block come from the same thread
  \item if $N\geq 1$, $\alpha = B_1\circ\cdots\circ B_{N}$
  \item every lock statement in $\alpha$ occurs as the first
    statement of some block $B$
  \item all statements in a block $B$ other than the first statement
    of $B$ are nsync statements
  \item for $1\leq i\leq N-1$: if $B_{i+1}$ is initial then $B_i$ is
    initial and $\tid(B_i)<\tid(B_{i+1})$
  \item if $B$ is a non-initial block and $u$ is the initial node of
    $B$, then no thread is normal at $u$ and
    $\ample_\graph(u)=\enabled(u)$, and
  \item if $B$ is a non-initial block, $u$ is any node in $B$, and
    $i\in\TID\setminus\{\tid(B)\}$ then thread $i$ is not normal at
    $u$.
  \end{enumerate}
\end{lemma}

\begin{proof}
  The first four follow immediately from the definition of
  \emph{block}.

  (5).  Assume $B_i$ is not initial; we will show $B_{i+1}$ is not
  initial.  Let $s$ be the initial state of $B_i$, and $s'$ the
  initial state of $B_{i+1}$.  Let $j=\tid(B_i)$.  As $B_i$ is not
  initial, thread $j$ is not normal at $s$.  Since the first
  statement of $B_i$ is in thread $j$, Definition \ref{def:race-space}
  implies that no thread is normal at $s$.  Since all statements in
  $B_i$ are in thread $j$, all threads $k$, for $k\neq j$, are in the
  same state at $s'$ that they were in $s$, so thread $k$ is not
  normal at $s'$.  But thread $j$ also cannot be normal at $s'$,
  else the block $B_i$ would not end at $s'$.  Hence no thread is
  normal at $s'$; in particular, $B_{i+1}$ is not initial.

  Now assume $B_{i+1}$---and therefore $B_i$---are initial.  Let $t$ be the last
  statement of $B_i$ and $t'$ the first statement of $B_{i+1}$.  Let
  $s$ be the state immediately preceding $t$ and $s'$ the state
  immediately following $t$ and preceding $t'$.  As $B_{i+1}$ is
  initial, thread $\tid(t')$ is normal at $s'$.  Thread $\tid(t)$ is
  normal at $s$.  (If $t$ is the first transition of $B_i$, this
  follows because $B_i$ is initial.  If $t$ is not the first
  transition of $B_i$, then this follows from the definition of
  \emph{block}.)  Now $\tid(t')\neq \tid(t)$, else the two transitions
  would be in the same block.  As $t$ is an nsync statement, it cannot
  enable or disable transitions in other threads, so $t'$ is also
  enabled at $s$, and thread $\tid(t')$ is also normal at $s$, as the
  local state of thread $\tid(t')$ is the same at $s$ or $s'$.  By
  Definition \ref{def:race-space}, $\tid(t)<\tid(t')$.  This means
  $\tid(B_i)<\tid(B_{i+1})$.

  (6). If some thread were normal at $u$, then the ample set for $u$
  would consist of the enabled statements from one of the normal
  threads, so the thread of the first transition of $B$ would be
  normal at $u$, i.e., $B$ would be initial.  Since $B$ is not
  initial, no thread is normal at $u$, and therefore $u$ is fully
  enabled in $G$.

  (7). By (6), at the initial state of $B$, no thread is normal.  All
  statements in $B$ belong to thread $\tid(B)$, and these statements
  cannot enable an nsync statement in another thread.  All threads
  other than thread $\tid(B)$ remain at their original local states at
  all nodes in $B$.  It follows that these threads are not normal at
  any node in $B$. \qed
\end{proof}

From Lemma \ref{lem:block-omnibus}, we conclude that the block
decomposition of a path $\alpha$ in $\graph$ has a prefix of initial
blocks in which the block $\tid$ is strictly increasing.  In
particular, no two initial blocks have the same $\tid$.  This is
followed by a sequence of non-initial blocks, each of which starts
with a statement in a non-normal thread $p$, and is followed by some
number of nsync statements in $p$.  Each of these nsync statements
is executed from a state at which thread $p$, and only thread $p$,
is normal.

\begin{definition}
  \label{def:complete}
  Let $\alpha$ be a finite path in $\graph$ and $B$ a block of
  $\alpha$.  We say $B$ is \emph{complete} if thread $\tid(B)$ is not
  normal at the final node of $B$.  We say $\alpha$ is
  \emph{complete} if every block of $\alpha$ is complete.  \qed
\end{definition}

Note: a path of length $0$ has $0$ blocks and is vacuously complete.

\begin{lemma}
  \label{lem:block-complete}
  Let $\alpha$ be a finite path in $\graph$. Then every block of
  $\alpha$ other than the last is complete.
\end{lemma}
\begin{proof}
  Let $B$ be a block of $\alpha$ that is not the last block,
  $i=\tid(B)$, and let $u\stackrel{t}{\ra}v$ be the last edge of $B$,
  and $v\stackrel{t'}{\ra}w$ the first edge of the next block $B'$.
  
  Assume thread $i$ is normal at $v$; we will arrive at a
  contradiction.  By Definition \ref{def:race-space}, $t'$ must belong
  to a thread that is normal at $v$, so $t'$ is an nsync statement.
  Let $j=\tid(t')$.  We have $i\neq j$, else $B'$ and $B$ would form a
  single block.  Now $t'$ must be enabled at $u$, since no statement
  from another thread can enable an nsync statement.  Moreover, $j$ is
  normal at $u$, since the local state of thread $j$ is the same at
  $u$ and $v$.  By Definition \ref{def:race-space}, $i<j$.  But since
  $i$ is normal at $v$, there is some nsync statement in thread $i$
  enabled at $\state(v)$, whence $j<i$, a contradiction. \qed
\end{proof}

Hence, only the last block of $\alpha$ could be incomplete.
However, we now show that under reasonable assumptions, $\alpha$
can be extended to a complete path.

\begin{lemma}
  \label{lem:complete}
  Let $u_0\in V$ and assume the set of nodes in $\graph$ reachable
  from $u_0$ is finite.  Then every finite path starting from $u_0$
  can be extended to a complete path.
\end{lemma}

\begin{proof}
  Say
  $\alpha=(u_0\stackrel{t_1}{\ra}\cdots\stackrel{t_{n}}{\ra}u_{n})$.
  If $n=0$, $\alpha$ is vacuously complete, so assume $n\geq 1$.

  By Lemma \ref{lem:block-complete}, every block other than the last
  is complete.  Suppose the last block is not complete.  Let
  $i=\tid(t_n)$.  Thread $i$ is normal at $u_n$.  Hence some (nsync)
  statement $t$ in thread $i$ is enabled at $u_n$.  Moreover, $i$ must
  be the least ID of a normal thread at $u_n$, because if there were
  some other normal thread $j$ at $u_n$, with $j<i$, then thread $j$
  would also be normal at $u_{n-1}$, contradicting the assumption
  that $\alpha$ is a path in $\graph$.  Hence
  $t\in\ample_{\graph}(u_n)$.  Append $t$, and the resulting node, to
  $\alpha$, and the result is still a path in $G$.

  Repeat the above as long as the path is not complete.  We claim that
  eventually, the path must become complete.  Otherwise, there is an
  infinite path in $G$, starting from a node reachable from $u_0$, in
  which thread $i$ is normal at every node, contradicting Lemma
  \ref{lem:cycle}. \qed
\end{proof}

Note: if a finite execution has a data race, then any extension will
also have a data race, by Lemma \ref{lem:extend-race}.

\begin{definition}
  \label{def:block-event}
  Let
  $\alpha=(u_0\stackrel{t_1}{\ra}\cdots\stackrel{t_{n}}{\ra}u_{n})$ be
  a finite path in $\graph$ with
  $\tr(\overline{\alpha})=e_1\ldots e_{n}$.  Define $N$,
  $i_0, \ldots, i_{N}$, and $B_1,\ldots,B_{N}$ as in Definition
  \ref{def:block}. For $1\leq j\leq N$, let
  \[
    b_j = e_{i_{j-1}+1}\cdots e_{i_j}.
  \]
  The sequence $b_j$ is the \emph{event sequence} of block $B_j$; it
  is a sequence of elements of $[\overline{\alpha}]$.  The sequence
  $b_1\cdots b_{N}$ is the \emph{block-event string} of $\alpha$; it
  is a sequence of event sequences.
  \qed
\end{definition}

We will adopt the following notational shorthand.  Suppose
$b_1, \ldots, b_n$ are event sequences; say
\[
  b_j=\langle t_{j,1},n_{j,1} \rangle
  \langle t_{j,2},n_{j,2} \rangle
  \cdots
  \langle t_{j,m_j},n_{j,m_j}\rangle.
\]
We will write \emph{there is an execution of the form}
\[
  s_0\stackrel{b_1}{\ra}
  s_1\stackrel{b_2}{\ra}
  \cdots
  \stackrel{b_{n}}{\ra}
  s_n
\]
to mean there is an execution of the form
\[
  s_0\stackrel{t_{1,1}}{\ra}
  s_{0,1}\stackrel{t_{1,2}}{\ra}
  \cdots
  \stackrel{t_{1,m_1}}{\ra}
  s_1\stackrel{t_{2,1}}{\ra}
  s_{1,1}\stackrel{t_{2,2}}{\ra}
  \cdots
  \stackrel{t_{n,m_n}}{\ra}
  s_n.
\]
A similar notation will be used for paths in place of executions.

\begin{lemma}
  \label{lem:block-commute}
  Let $\graph$ be the race-detecting \stategraph{} of $P$.  Let
  $\alpha$ be a path in $\graph$ with initial node $v$, blocks
  $B_1,\ldots,B_{N}$, and block-event string $b_1\cdots b_{N}$.
  Suppose
  \begin{enumerate}
  \item $1\leq i<N$,
  \item $B_i$ is not an initial block,
  \item $B_{i+1}$ is complete,
  \item the first statement of $B_i$ does not happen before 
    the first statement of $B_{i+1}$, and
  \item no statement of $B_i$ conflicts with a statement of
    $B_{i+1}$.
  \end{enumerate}
  Then $b_1\cdots b_{i-1}b_{i+1}b_ib_{i+2}\cdots b_{N}$ is the
  block-event string of a path $\alpha'$ from $v$ in $G$ with
  $\DR(\overline{\alpha'})=\DR(\overline{\alpha})$.
\end{lemma}
\begin{proof}
  Let $v_j$ be the node in $\alpha$ just before $b_{j+1}$
  ($0\leq j<N$) and let $v_N$ be the final node of $\alpha$.
  Let $s_j=\state(v_j)$.  
  The execution $\overline{\alpha}$ has the form
  \[
    s_0\stackrel{b_1}{\ra} \cdots
    \stackrel{b_{i-1}}{\ra}
    s_{i-1}\stackrel{b_{i}}{\ra}
    s_i\stackrel{b_{i+1}}{\ra}
    s_{i+1}\stackrel{b_{i+2}}{\ra} \cdots
    \stackrel{b_{N}}{\ra}
    s_N.
  \]

  The idea is to transpose $b_i$ and $b_{i+1}$.  Since there are no
  conflicts, all nsync statements in one block commute with any
  statement in the other block, by Lemma \ref{lem:commute};
  furthermore these transpositions do not alter the data race
  relation.  The only statements which are possibly not nsync are the
  first statements of the two blocks, but assumption (4) and Lemma
  \ref{lem:commute} guarantee these commute, again without altering
  the data race relation.  Therefore there is an execution of the form
  \[
    s_0\stackrel{b_1}{\ra} \cdots
    \stackrel{b_{i-1}}{\ra}
    s_{i-1}\stackrel{b_{i+1}}{\ra}
    s'\stackrel{b_{i}}{\ra}
    s_{i+1}\stackrel{b_{i+2}}{\ra} \cdots
    \stackrel{b_{N}}{\ra}
    s_N,
  \]
  with data race relation $\DR(\overline{\alpha})$.  We show the
  execution corresponds to a path in \graph.

  Let $p=\tid(B_i)$ and $q=\tid(B_{i+1})$.  By assumption (4),
  $p\neq q$.  Since $B_i$ is not initial, by Lemma
  \ref{lem:block-omnibus}(6), no thread is normal at $s_{i-1}$, and
  $\ample_\graph(v_{i-1})=\enabled(v_{i-1})$.  Hence the first
  statement $t_1$ of $b_{i+1}$ is in $\ample_\graph(v_{i-1})$.  Let
  $\tilde{s}_1=\execute(s_{i-1}, t_1)$.  Thus there is a node
  $\tilde{v}_1\in V$ with $\state(\tilde{v}_1)=\tilde{s}_1$, and an
  edge $v_{i-1}\stackrel{t_1}{\ra}\tilde{v}_1\in E$.

  Now the local state of thread $q$ is the same at $s_{i-1}$ and
  $s_{i}$.  Hence the local state of $q$ at $\tilde{s}_1$ is the same
  as the local state of $q$ at $\execute(s_i,t_1)$.  In particular, if
  there is a second statement $t_2$ in $B_{i+1}$, then $t_2$ is a sync
  statement that is also enabled at $\tilde{s}_1$ and $q$ is the only
  thread that is normal at $\tilde{s}_1$.  Let
  $\tilde{s}_2=\execute(\tilde{s}_1,t_2)$.  Hence there is a node
  $\tilde{v}_2\in V$ with $\state(\tilde{v}_2)=\tilde{s}_2$ and an
  edge $\tilde{v}_1\stackrel{t_2}{\ra}\tilde{v}_2 \in E$.
  
  Continuing in this way, we see that $b_{i+1}$ defines a path in
  $\graph$ from $v_{i-1}$ to some $v'\in V$ with $\state(v')=s'$.
  Furthermore, $q$ is not normal at $s'$.

  Since no thread other than $q$ has changed its local state in the
  path from $v_i$ to $v'$ defined by $b_{i+1}$, at $s'$ no thread is
  normal.  So again $\ample_\graph(v')=\enabled(v')$ and $b_i$
  defines a path in $\graph$ from $v'$ to a node $v'_{i+1}$ with
  $\state(v'_{i+1})=s_{i+1}$.

  We now have two paths in $\graph$, both starting at $v_i$.  The
  first executes $b_ib_{i+1}$ and ends at $v_{i+1}$; the second
  executes $b_{i+1}b_{i}$ and ends at $v'_{i+1}$.  We claim
  $v'_{i+1}=v_{i+1}$.  We already know both nodes have the same state
  component, $s_{i+1}$; we must show they have the same
  $\ma$-component.  But $\ma_p$ is determined solely by the sequence
  of statements in thread $p$, and $\ma_q$ solely by those of thread
  $q$.  Since $p\neq q$, the $\ma$-component is the same after
  $b_ib_{i+1}$ or $b_{i+1}b_{i}$.

  Hence the string $b_1\cdots b_{i-1}b_{i+1}b_ib_{i+2}\cdots b_{N}$
  defines a path $\alpha'$ in $G$ from $v$.  We just need to see this
  string is the block-event string of $\alpha'$.

  Clearly, the paths $v_0\stackrel{b_1}{\ra}v_1$, \ldots,
  $v_{i-2}\stackrel{b_{i-1}}{\ra}v_{i-1}$ are the first $i-1$
  blocks of $\alpha'$, as they are the first $i-1$ blocks
  of $\alpha$.  Consider the remaining paths
  \[
    v_{i-1}\stackrel{b_{i+1}}{\ra}v', \:\:
    v'\stackrel{b_{i}}{\ra}v_{i+1}, \:\:
    v_{i+1}\stackrel{b_{i+2}}{\ra}v_{i+2}, \:\: \cdots, \:\:
    v_{N-1}\stackrel{b_{N}}{\ra}v_N.
  \]
  Since $B_i$ is not initial, by Lemma \ref{lem:block-omnibus}(6), no
  thread is normal at $v_{i-1}$.  We have already seen that no thread
  is normal at $s'=\state(v')$.  By Lemma \ref{lem:block-omnibus}(5),
  $B_j$ is not initial for $j\geq i$.  Hence no thread is normal at
  $v_{i+1},\ldots,v_{N-1}$.  It follows that these paths are the
  remaining $N-i+1$ blocks of $\alpha'$. \qed
\end{proof}

\subsection{Blocks and data races}

Let $\alpha$ be a finite path in $\graph$.  We say two blocks $B$ and
$B'$ of $\alpha$ \emph{race} if there is an event $e$ in $B$ and $e'$
in $B'$ such that $e$ and $e'$ race in $\alpha$.  We say an event $e$
in $\alpha$ \emph{happens before} $B$ if $e$ happens before the first
event of $B$.  We say $B$ \emph{happens before} $B'$ if the first
event in $B$ happens before the first event in $B'$.

\begin{definition}
  \label{def:distance}
  Let $\alpha$ be a finite path in $\graph$ with blocks
  $B_1,\ldots,B_{N}$.  Suppose $B_{N}$ races with a prior block.  Let
  $i$ be the maximum integer in $[1,N-1]$ such that $B_i$ races with
  $B_{N}$.  Let $j$ be the maximum integer in $[1,N]$ such that
  $\tid(B_j)=\tid(B_i)$.  (Note $i\leq j<N$.)  The \emph{race
    distance} of $\alpha$ is $N-j$.  \qed
\end{definition}
Hence the race distance is the number of blocks that occur after the
last block from the last thread that races with $B_{N}$.  Note the
race distance is at least 1.

\begin{lemma}
  \label{lem:race-structure}
  Let $u_0\in V$. Assume the set of nodes reachable from $u_0$ in $G$
  is finite. If $\graph$ has a path starting from $u_0$ with a data
  race, then there is a finite complete path $\beta$ in $\graph$
  starting from $u_0$, with blocks $B_1,\ldots B_{N}$ satisfying the
  following: there is some $i\in[1,N-1]$ such that
  \begin{itemize}
  \item $B_i$ and $B_{N}$ race, and
  \item for all $j\in[i+1,N]$, $\tid(B_j)\neq\tid(B_i)$.
  \end{itemize}
\end{lemma}

\begin{proof}
  If $\graph$ has a path starting from $u_0$ with a data race, then
  truncate the path after the second event involved in the race, and
  the resulting finite path also has a data race.  So we may assume
  that $\graph$ has a finite path with a data race.
  
  By Lemma \ref{lem:complete}, any finite path can be extended to a
  complete path, and if the original contained a data race, so will
  the extension, by Lemma \ref{lem:extend-race}.  Hence we may assume
  $\graph$ contains a complete path starting from $u_0$ with a race.
  
  Let $N$ be the minimal block length of any complete path starting
  from $u_0$ with a data race.  All complete paths of block length $N$
  containing data races must have the last block racing with a
  previous block, else there would be a complete path of smaller block
  length with a data race.  Among all such paths, let $\alpha$ be one
  with minimal race distance.  Let $B_1,\ldots,B_{N}$ be the blocks of
  $\alpha$.

  Hence there is an event in $\alpha$ occurring before $B_{N}$ that
  races with some event in $B_{N}$.  Let $t_1$ be the last such event.
  Say $t_1$ occurs in block $B_i$.
  % The race distance of $\alpha$ is $N-i$.
  Let $p=\tid(B_i)$.  We have
  \begin{enumerate}
  \item There is no racy path from $u_0$ with block length less 
    than $N$. 
  \item There is no racy path from $u_0$ of block length $N$
    with race distance less than that of $\alpha$.  %$N-i$. 
  \item Any two events occurring before $B_{N}$ that have conflicting
    transitions are ordered by happens-before. (Else there would be a
    racy path with block length strictly less than $N$.)
  \item No event in $B_{i+1},\ldots,B_{N-1}$ races with any event in
    $B_{N}$.  (As $t_1$ is the last event to race with one in
    $B_{N}$.)
  \end{enumerate}

  Suppose a block from thread $p$ occurs after $B_i$.  We will
  arrive at a contradiction.

  Let $B_j$ be the last block from thread $p$.  By assumption,
  $i<j<N$.  According to Definition \ref{def:distance}, the race
  distance of $\alpha$ is $N-j$.  Note $B_j$ is not initial, since
  there is a previous block from the same thread, and if both were
  initial it would contradict Lemma \ref{lem:block-omnibus}(5).  Let
  $t_2$ be the first event of $B_j$.

  Is there a block $B$ between $B_j$ and $B_{N}$ such that $t_2$
  does not happen before $B$?  There are two cases, both of
  which lead to the desired contradiction.

  \textbf{Case 1:} there is some $k\in[j+1,N-1]$ such that $t_2$ does
  not happen before $B_k$.  Choose the minimal such $k$.  Then for
  $j<l<k$, $t_2$ happens before $B_l$, but $t_2$ does not happen
  before $B_k$.  The block-event string of $\alpha$ has the form
  \[
    b_1\cdots b_i\cdots b_j\cdots b_k\cdots b_{N}.
  \]

  Suppose $j\leq l<k$. Then:
  \begin{itemize}
  \item $B_l$ does not happen before $B_k$.  (If $j=l$, then $B_l$
    does not happen before $B_k$, as $t_2$ does not happen before
    $B_k$.  If $j<l$ then $B_l$ does not happen before $B_k$, else
    $t_2$ happens before $B_l$ happens before $B_k$.)
  \item $B_k$ does not happen before $B_l$. (Since $B_k$ occurs
    after $B_l$ in $\alpha$.)
  \item Hence $B_k$ and $B_l$ are not ordered by happens-before.
    By (3), no statement from $B_l$ conflicts with any statement
    in $B_k$.
  \end{itemize}
  By Lemma \ref{lem:block-commute}, we may repeatedly transpose $b_k$
  with the block to its left, until $b_k$ occurs just before $b_j$,
  i.e., there exists a path in $\graph$ from $u_0$ with block-event
  string
  \[
    b_1\cdots b_i\cdots b_{j-1}b_kb_j\cdots b_{k-1}b_{k+1}\cdots b_{N}.
  \]
  This path has a data race, has block length $N$, but has race
  distance $N-j-1$, one less than that of $\alpha$, contradicting (2).

  \textbf{Case 2:} for all $k\in[j+1,N-1]$, $t_2$ happens before
  $B_k$.

  We claim: for $j\leq k<N$, $B_k$ does not happen before $B_{N}$.
  (Proof: we know $B_j$ does not happen before $B_{N}$, else $t_1$
  happens before $B_j$ happens before $B_{N}$, contradicting the
  assumption that $t_1$ races with $B_{N}$.  If $j<k<N$, $B_k$ does
  not happen before $B_{N}$, else $t_2$ happens before $B_k$ happens
  before $B_{N}$, i.e., $B_j$ happens before $B_{N}$.)

  It follows from (4) that for $j\leq k<N$, $B_k$ contains no
  statement that conflicts with one in $B_{N}$.

  As $j\leq N-1<N$, $B_{N-1}$ does not conflict with $B_{N}$ and
  $B_{N-1}$ does not happen before $B_{N}$.  By Lemma
  \ref{lem:block-commute}, there is a path in $\graph$ from $u_0$ with
  block-event string
  \[
    b_1\cdots b_i\cdots b_j\cdots b_{N-2}b_{N}b_{N-1}
  \]
  and with a data race occurring between an event in $b_i$ and one in
  $b_{N}$.  Truncate the path just after $b_{N}$, and the resulting
  path has block length $N-1$ and contains a data race, contradicting
  (1) (the minimality of $N$). \qed
\end{proof}

% Note: you can actually show that if B_i is not initial, then
% you can make i=N-1, i.e., the two racing blocks are adjacent.
% But if B_i is initial, other blocks may intervene.  Example:
%
% thread 1: x=1;
% thread 2: acq(l); y=1; rel(l);
% thread 3: acq(l); if (y==1) x=2; rel(l).
% This has a racy execution:
%  t1:[x=1]  t2:[acl(l);y=1]  t2:[rel(l)]  t3:[acq(l); x=2] t3:[rel(l)].
% You can't move blocks B2 and B3 to the left of B1 because B1 is
% initial.

\subsection{Proof of Theorem \ref{thm:detect} for Programs Without
  Barriers}
\label{sec:proofnobar}

We can now complete the proof of Theorem \ref{thm:detect} in the case
where $P$ has no barriers.  By Lemma \ref{lem:dense}, $\graph$ is
dense.  Part 2 of Theorem \ref{thm:detect} then follows from part 2 of
Theorem \ref{thm:reduce}.

We now turn to the proof of part 1 of Theorem \ref{thm:detect}.  As
Lemma \ref{lem:sound} proves one direction, we must prove the other
direction.  So suppose there is an execution from $s_0$ with a data
race.  We must show there is a path in $\graph$ from $u_0$ which
detects a race.

By Theorem \ref{thm:reduce}(1), there is a path in $\graph$ from $u_0$
with a data race.  By Lemma \ref{lem:race-structure}, there is a
complete path $\alpha$ in $\graph$ from $u_0$, with blocks
$B_1,\ldots,B_N$, and $i\in[1,N-1]$, such that $B_i$ races with $B_N$
and $B_i$ is the last block from thread $p=\tid(B_i)$ in $\alpha$.
Let $t_1$ be the event in $B_i$ and $t_2$ the conflicting event in
$B_N$.  At the end of $B_i$, $t_1$ is in $\ma_p$.  As thread $p$ does
not execute again, $t_1$ remains in $\ma_p$ at the final node $v$ of
$\alpha$.  Moreover, $t_2$ is in $\ma_q$ at $v$, where $q=\tid(B_N)$.
The last transition of $B_N$ brings thread $q$ to an $R_q$ or terminal
state, and hence detects a data race involving $t_1$ and $t_2$.

% By Lemma \ref{lem:cycle}, there is a finite path from $v$ in $\graph$
% ending at a final node (i.e., a node with no outgoing edge), or at a
% node at which no thread is normal.  Let $w$ be the first node on this
% path at which at least one of the following holds: the local state of
% $p$ is in $\LockingState_p$ at $w$, the local state of $q$ is in
% $\LockingState_q$ at $w$, no thread is normal at $w$, or $w$ is a
% final node.  Let $\beta$ be the prefix of this path ending at $w$.

% Suppose $e=u\stackrel{t}{\ra}u'$ is an edge in $\beta$ with
% $\tid(t)=p$.  Then the local state $\sigma$ of $p$ at $u$ is not in
% $R_p$.  For such an edge can occur in $\graph$ only when no thread is
% normal at $u$, but $w$ was chosen so that no previous node in $\beta$
% had this property.  Since $\ma_p$ is emptied only when executing an
% edge $e$ for $\tid(t)=p$ and $\sigma\in R_p$, we conclude $t_1$ is in
% $\ma_p$ at $w$.  By the same reasoning, $t_2$ is in $\ma_q$ at $w$.

% If the local state of $p$ is in $\LockingState_p$ at $w$, then the
% last edge of $\beta$ detects a race.  Similarly, if the local state of
% $q$ is in $\LockingState_q$ at $w$, the last edge of $\beta$ detects a
% race.  Suppose no thread is normal at $w$, and the local state of
% thread $p$ is not in $\LockingState_p$ at $w$.  Then $w$ is fully
% enabled, and the local state of $p$ is in
% $R_p\setminus\LockingState_p$ at $w$.  So there is an edge in $\graph$
% from $w$ on a transition $t$ in $p$, and this edge detects a race.
% Finally, if $w$ is final, thread $p$ (or $q$) detects a race at $w$.

\subsection{Proof of Theorem \ref{thm:detect}: General Case}

The general case of the theorem can be reduced to the case of a
program with no barriers in a simple way.  The key observation is that
if a race occurs in an execution of a program with barriers, then the
two conflicting events must occur in the same barrier epoch.  This is
because any two events in different epochs are ordered by
happens-before, and thus cannot form a data race.

The reduction to the barrier-free case requires a simple program
transformation.  Given a program $P$ (that may contain barriers) let
$P'$ be the program that is the same in every respect as $P$, except
that every barrier state in $P$ is made a terminal state in $P'$.

Suppose $\zeta$ is an execution of $P$, $i\geq 0$, and there is at
least one transition in barrier epoch $i$ in $\zeta$.  Define an
execution $\zeta'$ of $P'$ that extracts epoch $i$ of $\zeta$ as
follows.  The initial state $s_0'$ of $\zeta'$ is defined as follows:
if $i=0$, $s_0'$ is the initial state of $\zeta$.  Otherwise, the lock
state of $s_0'$ is the lock state in the state of $\zeta$ when all
threads are in the $i$-th barrier.  The thread state of thread $p$ in
$s_0'$ is the thread state of $p$ just after the $i$-th
$\barrierExit_p$ transition in $\zeta$; if thread $p$ does not execute
$i$ $\barrierExit_p$ transitions in $\zeta$ (i.e., $p$ never enter
epoch $i$), then the thread state of $p$ in $s_0'$ is the final thread
state of $p$ in $\zeta$.  The transitions in $\zeta'$ are precisely
the transitions in epoch $i$ of $\zeta$, excluding the initial
sequence of barrier-exit transitions in epoch $i$.  Clearly, if
$\zeta$ has a data race that occurs in epoch $i$, then $\zeta'$ has a
data race.

Let $G$ be the race-detecting state graph of $P$, and $G'$ the
race-detecting state graph of $P'$.  Suppose now that $\alpha$ is a
path in $G$ with at least one transition in epoch $i$.  Let
$\zeta=\overline{\alpha}$.  There is a corresponding path $\alpha'$ in
$G'$, specified as follows: the initial node of $\alpha'$ is
$\langle s_0',\emptyset^{\TID}\rangle$ and
$\overline{\alpha'}=\zeta'$.

Now suppose $P$ has an execution $\zeta$ with a data race.  Let $s_0$
be the initial state of $\zeta$.  Say the first data race in $\zeta$
occurs in epoch $m\geq 0$.  We will construct a path $\alpha$ in $G$
that starts at $v_0=\langle s_0,\emptyset^{\TID}\rangle$ and
detects a race.

Suppose $0\leq i<m$.  Consider the execution $\zeta'_i$ of $P'$
obtained by extracting epoch $i$ from $\zeta$.  Let $s'$ be the
initial state of $\zeta'_i$ and $s''$ the final state.  Since $i$ is
not the last barrier epoch of $\zeta$, every thread must be terminal
at $s''$.  There is no data race in $\zeta'_i$, because the first race
in $\zeta$ occurs in epoch $m$. By Theorem \ref{thm:reduce}(2), there
is a path $\alpha'_i$ in $G'$ from
$\langle s',\emptyset^{\TID}\rangle$ to a node $v''$ in $G'$ with
state component $s''$.

The paths $\alpha'_0,\ldots,\alpha'_{m-1}$ can be ``stitched
together'' to form a path $\beta$ in $G$ as follows: for $1\leq i<m$
and $j\in\TID$, insert statement $\barrierExit_j$ just before the
first transition from thread $j$ in $\alpha'_i$.  The path $\beta$
terminates at a node whose state component is the state of $\zeta$ at
the beginning of epoch $m$.

Now consider the execution $\zeta'_m$ of $P'$ obtained by extracting
epoch $m$ of $\zeta$.  This execution has a data race.  We may apply
Theorem \ref{thm:detect} to $\zeta'_m$, since $P'$ has no barriers.
Thus there exists a path in $P'$ from a node whose initial state
component is the initial state of $\zeta'_m$, and which detects a
race.  Stitch this path onto $\beta$ to yield a path in $G$ which
detects a race.  This proves part 1.

The proof of part 2 is almost exactly the same.  Given an execution
ending at a final state, again break it up into epochs and apply the
barrier-free version of the theorem to the $P'_i$.  Stitch the
resulting paths together to yield a path in $G$ ending at a node with
state component the final state.

\section{Full Results}

This section shows the expected result (data race or no data race),
and the results reported by CIVL and LLOV for all test cases.  It also
shows runtimes for CIVL on an M1 MacBook Pro with 16GB memory. The
additional test cases are shown in Table~\ref{tbl:newcases}, and the
DataRaceBench test cases are shown in Tables~\ref{tbl:drbone}
and~\ref{tbl:drbtwo}.  See Section~\ref{sec:eval} for a description of
the modifications (including imposition of bounds on inputs and thread
counts) made to the DataRaceBench programs for CIVL.

\begin{table}[]
\begin{adjustbox}{width=\textwidth}
\begin{tabular}{p{6.7cm}p{1.2cm}p{1.2cm}p{1.2cm}p{1.2cm}}
\textbf{Filename}    & \textbf{CIVL\newline time} & \textbf{Expec.\newline Result} & \textbf{CIVL\newline Result}                         & \textbf{LLOV\newline Result} \\ \hline 
sync1\_no.c                           & \phantom{0}1.09                                   & N                                     & N                                        & \cellcolor[HTML]{F4CCCC}P                \\
sync1\_yes.c                          & \phantom{0}1.31                                   & P                                     & P                                        & P                                        \\
critsec3\_no.c                        & \phantom{0}1.00                                   & N                                     & N                                        & \cellcolor[HTML]{F4CCCC}P                \\
critsec3\_yes.c                       & \phantom{0}1.00                                   & P                                     & P                                        & P                                        \\
atomic3\_no.c                         & \phantom{0}1.02                                   & N                                     & N                                        & \cellcolor[HTML]{F4CCCC}P                \\
atomic3\_yes.c                        & \phantom{0}1.04                                   & P                                     & P                                        & P                                        \\
bar1\_no.c                            & \phantom{0}7.19                                   & N                                     & N                                        & \cellcolor[HTML]{F4CCCC}P                \\
bar1\_yes.c                           & \phantom{0}1.24                                   & P                                     & P                                        & P                                        \\
bar2\_no.c                            & \phantom{0}1.17                                   & N                                     & N                                        & \cellcolor[HTML]{F4CCCC}P                \\
bar2\_yes.c                           & \phantom{0}1.17                                   & P                                     & P                                        & P                                        \\
bar3\_no.c                            & \phantom{0}3.56                                   & N                                     & N                                        & \cellcolor[HTML]{F4CCCC}P                \\
bar3\_yes.c                           & \phantom{0}1.43                                   & P                                     & P                                        & P                                        \\
diffusion1\_no.c                      & \phantom{0}3.12                                   & N                                     & N                                        & N                                        \\
diffusion1\_yes.c                     & \phantom{0}1.38                                   & P                                     & P                                        & \cellcolor[HTML]{F4CCCC}N                \\
diffusion2\_no.c                      & 22.19                                             & N                                     & N                                        &  \cellcolor[HTML]{F4CCCC}-                                     \\
diffusion2\_yes.c                     & \phantom{0}5.05                                   & P                                     & P                                        &  \cellcolor[HTML]{F4CCCC}-                                     \\
critsec2\_no.c                        & \phantom{0}2.02                                   & N                                     & N                                        &  \cellcolor[HTML]{F4CCCC}-                                     \\
critsec2\_yes.c                       & \phantom{0}1.36                                   & P                                     & P                                        &  \cellcolor[HTML]{F4CCCC}-                                     \\
prodcons\_no.c                        & 22.60                                             & N                                     & N                                        &  \cellcolor[HTML]{F4CCCC}-                                     \\
prodcons\_yes.c                       & \phantom{0}1.33                                   & P                                     & P                                        &  \cellcolor[HTML]{F4CCCC}-                                    \\ \hline
\end{tabular}
\end{adjustbox}
\vspace{0.1cm}
\caption{Results of additional test cases. CIVL runtime in seconds; Expected Result: P (positive) = data race detected, N (negative) = no race detected; CIVL Result; LLOV Result}
\label{tbl:newcases}
\end{table}

\begin{table}
\centering
\begin{adjustbox}{width=\textwidth}
\begin{tabular}{p{6.7cm}p{1.2cm}p{1.2cm}p{1.2cm}p{1.2cm}}
\textbf{Filename}    & \textbf{CIVL\newline time} & \textbf{Expec.\newline Result} & \textbf{CIVL\newline Result}                         & \textbf{LLOV\newline Result} \\ \hline 
DRB001-antidep1-orig-yes.c               & \phantom{00}1.44                                   & P                                     & P                                                                & P                                        \\
DRB002-antidep1-var-yes.c                & \phantom{00}1.53                                   & P                                     & P                                                                & P                                        \\
DRB003-antidep2-orig-yes.c               & \phantom{00}1.51                                   & P                                     & P                                                                & P                                        \\
DRB004-antidep2-var-yes.c                & \phantom{00}1.69                                   & P                                     & P                                                                & P                                        \\
DRB005-indirectaccess1-orig-yes.c        & \phantom{00}2.19                                   & P                                     & P                                                                & P                                        \\
DRB006-indirectaccess2-orig-yes.c        & \phantom{00}2.35                                   & P                                     & P                                                                & P                                        \\
DRB007-indirectaccess3-orig-yes.c        & \phantom{00}2.34                                   & P                                     & P                                                                & P                                        \\
DRB008-indirectaccess4-orig-yes.c        & \phantom{00}2.11                                   & P                                     & P                                                                & P                                        \\
DRB009-lastprivatemissing-orig-yes.c     & \phantom{00}1.31                                   & P                                     & P                                                                & P                                        \\
DRB010-lastprivatemissing-var-yes.c      & \phantom{00}1.60                                   & P                                     & P                                                                & P                                        \\
DRB011-minusminus-orig-yes.c             & \phantom{00}1.57                                   & P                                     & \cellcolor[HTML]{FFFFFF}P                                        & P                                        \\
DRB012-minusminus-var-yes.c              & \phantom{00}1.57                                   & P                                     & \cellcolor[HTML]{FFFFFF}P                                        & P                                        \\
DRB013-nowait-orig-yes.c                 & \phantom{00}1.39                                   & P                                     & \cellcolor[HTML]{FFFFFF}P                                        & P                                        \\
DRB014-outofbounds-orig-yes.c            & \phantom{00}1.15                                   & P                                     & \cellcolor[HTML]{F4CCCC}N                                        & P                                        \\
DRB015-outofbounds-var-yes.c             & \phantom{00}1.44                                   & P                                     & \cellcolor[HTML]{F4CCCC}N                                        & P                                        \\
DRB016-outputdep-orig-yes.c              & \phantom{00}1.30                                   & P                                     & \cellcolor[HTML]{FFFFFF}P                                        & P                                        \\
DRB017-outputdep-var-yes.c               & \phantom{00}1.66                                   & P                                     & \cellcolor[HTML]{FFFFFF}P                                        & P                                        \\
DRB018-plusplus-orig-yes.c               & \phantom{00}1.55                                   & P                                     & \cellcolor[HTML]{FFFFFF}P                                        & P                                        \\
DRB019-plusplus-var-yes.c                & \phantom{00}1.84                                   & P                                     & \cellcolor[HTML]{FFFFFF}P                                        & P                                        \\
DRB020-privatemissing-var-yes.c          & \phantom{00}1.69                                   & P                                     & \cellcolor[HTML]{FFFFFF}P                                        & P                                        \\
DRB021-reductionmissing-orig-yes.c       & \phantom{00}1.46                                   & P                                     & P                                                                & P                                        \\
DRB022-reductionmissing-var-yes.c        & \phantom{00}1.65                                   & P                                     & P                                                                & P                                        \\
DRB023-sections1-orig-yes.c              & \phantom{00}1.28                                   & P                                     & P                                                                & P                                        \\
DRB028-privatemissing-orig-yes.c         & \phantom{00}1.47                                   & P                                     & \cellcolor[HTML]{FFFFFF}P                                        & P                                        \\
DRB029-truedep1-orig-yes.c               & \phantom{00}1.51                                   & P                                     & \cellcolor[HTML]{FFFFFF}P                                        & P                                        \\
DRB030-truedep1-var-yes.c                & \phantom{00}1.59                                   & P                                     & \cellcolor[HTML]{FFFFFF}P                                        & P                                        \\
DRB031-truedepfirstdimension-orig-yes.c  & \phantom{00}1.69                                   & P                                     & \cellcolor[HTML]{FFFFFF}P                                        & P                                        \\
DRB032-truedepfirstdimension-var-yes.c   & \phantom{00}2.15                                   & P                                     & \cellcolor[HTML]{FFFFFF}P                                        & P                                        \\
DRB033-truedeplinear-orig-yes.c          & \phantom{00}1.49                                   & P                                     & \cellcolor[HTML]{FFFFFF}P                                        & P                                        \\
DRB034-truedeplinear-var-yes.c           & \phantom{00}1.58                                   & P                                     & \cellcolor[HTML]{FFFFFF}P                                        & P                                        \\
DRB035-truedepscalar-orig-yes.c          & \phantom{00}1.45                                   & P                                     & P                                                                & P                                        \\
DRB036-truedepscalar-var-yes.c           & \phantom{00}1.60                                   & P                                     & P                                                                & P                                        \\
DRB037-truedepseconddimension-orig-yes.c & \phantom{0}15.89                                   & P                                     & \cellcolor[HTML]{FFFFFF}P                                        & P                                        \\
DRB038-truedepseconddimension-var-yes.c  & \phantom{00}2.05                                   & P                                     & \cellcolor[HTML]{FFFFFF}P                                        & P                                        \\
DRB039-truedepsingleelement-orig-yes.c   & \phantom{00}1.49                                   & P                                     & P                                                                & P                                        \\
DRB040-truedepsingleelement-var-yes.c    & \phantom{00}1.58                                   & P                                     & P                                                                & P                                        \\
DRB041-3mm-parallel-no.c                 & \phantom{0}85.10                                   & N                                     & N                                                                & N                                        \\
DRB043-adi-parallel-no.c                 & \phantom{}132.75                                   & N                                     & N                                                                & N                                        \\
DRB045-doall1-orig-no.c                  & \phantom{00}3.88                                   & N                                     & N                                                                & N                                        \\
DRB046-doall2-orig-no.c                  & \phantom{00}5.66                                   & N                                     & N                                                                & N                                        \\
DRB047-doallchar-orig-no.c               & \phantom{00}2.17                                   & N                                     & N                                                                & N                                        \\
DRB048-firstprivate-orig-no.c            & \phantom{00}4.09                                   & N                                     & N                                                                & N                                        \\ \hline
\end{tabular}
\end{adjustbox}
\vspace{0.1cm}
\caption{DataRaceBench results, part 1.}
\label{tbl:drbone}
\end{table}

\begin{table}
\centering
\begin{adjustbox}{width=\textwidth}
\begin{tabular}{p{6.7cm}p{1.2cm}p{1.2cm}p{1.2cm}p{1.2cm}}
\textbf{Filename}    & \textbf{CIVL\newline time} & \textbf{Expec.\newline Result} & \textbf{CIVL\newline Result}                         & \textbf{LLOV\newline Result} \\ \hline 
DRB050-functionparameter-orig-no.c       & \phantom{00}3.97                                    & N                                     & N                                                                & N                                        \\
DRB051-getthreadnum-orig-no.c            & \phantom{00}1.78                                    & N                                     & N                                                                & N                                        \\
DRB052-indirectaccesssharebase-orig-no.c & \phantom{00}2.83                                    & N                                     & N                                                                & \cellcolor[HTML]{F4CCCC}P                \\
DRB053-inneronly1-orig-no.c              & \phantom{00}4.20                                    & N                                     & N                                                                & N                                        \\
DRB054-inneronly2-orig-no.c              & \phantom{0}12.78                                   & N                                     & N                                                                & \cellcolor[HTML]{F4CCCC}P                \\
DRB055-jacobi2d-parallel-no.c            & \phantom{0}71.78                                   & N                                     & N                                                                & N                                        \\
DRB057-jacobiinitialize-orig-no.c        & \phantom{0}14.50                                   & N                                     & N                                                                & N                                        \\
DRB058-jacobikernel-orig-no.c            & \phantom{00}6.31                                   & N                                     & N                                                                & N                                        \\
DRB059-lastprivate-orig-no.c             & \phantom{00}4.41                                    & N                                     & N                                                                & N                                        \\
DRB060-matrixmultiply-orig-no.c          & \phantom{0}13.98                                   & N                                     & N                                                                & N                                        \\
DRB061-matrixvector1-orig-no.c           & \phantom{00}5.75                                    & N                                     & N                                                                & N                                        \\
DRB062-matrixvector2-orig-no.c           & \phantom{}156.76                        & N                                     & N                                                                & N                                        \\
DRB063-outeronly1-orig-no.c              & \phantom{00}5.36                                    & N                                     & N                                                                & N                                        \\
DRB064-outeronly2-orig-no.c              & \phantom{00}1.92                                    & N                                     & N                                                                & N                                        \\
DRB065-pireduction-orig-no.c             & \phantom{0}67.06                                   & N                                     & N                                                                & N                                        \\
DRB066-pointernoaliasing-orig-no.c       & \phantom{00}4.45                                    & N                                     & N                                                                & N                                        \\
DRB067-restrictpointer1-orig-no.c        & \phantom{00}4.45                                    & N                                     & N                                                                & N                                        \\
DRB068-restrictpointer2-orig-no.c        & \phantom{00}5.15                                    & N                                     & N                                                                & N                                        \\
DRB069-sectionslock1-orig-no.c           & \phantom{00}1.99                                    & N                                     & N                                                                & \cellcolor[HTML]{F4CCCC}P                \\
DRB073-doall2-orig-yes.c                 & \phantom{00}1.46                                    & P                                     & P                                                                & P                                        \\
DRB074-flush-orig-yes.c                  & \phantom{00}1.58                                    & P                                     & P                                                                & P                                        \\
DRB075-getthreadnum-orig-yes.c           & \phantom{00}1.18                                    & P                                     & p                                                                & P                                        \\
DRB076-flush-orig-no.c                   & \phantom{0}35.00                                   & N                                     & N                                                                & N                                        \\
DRB077-single-orig-no.c                  & \phantom{00}1.62                                    & N                                     & N                                                                & N                                        \\
DRB088-dynamic-storage-orig-yes.c        & \phantom{00}1.27                                    & P                                     & P                                                                & P                                        \\
DRB089-dynamic-storage2-orig-yes.c       & \phantom{00}1.22                                    & P                                     & P                                                                & P                                        \\
DRB090-static-local-orig-yes.c           & \phantom{00}2.09                                    & P                                     & P                                                                & P                                        \\
DRB093-doall2-collapse-orig-no.c         & \phantom{00}6.92                                    & N                                     & N                                                                & N                                        \\
DRB103-master-orig-no.c                  & \phantom{00}1.50                                    & N                                     & N                                                                & N                                        \\
DRB104-nowait-barrier-orig-no.c          & \phantom{00}5.26                                   & N                                     & N                                                                & N                                        \\
DRB108-atomic-orig-no.c                  & \phantom{00}8.66                                    & N                                     & N                                                                & N                                        \\
DRB109-orderedmissing-orig-yes.c         & \phantom{00}1.38                                    & P                                     & P                                                                & P                                        \\
DRB110-ordered-orig-no.c                 & \phantom{0}92.89                                   & N                                     & N                                                                & N                                        \\
DRB111-linearmissing-orig-yes.c          & \phantom{00}1.34                                    & P                                     & P                                                                & P                                        \\
DRB113-default-orig-no.c                 & \phantom{0}12.01                                   & N                                     & N                                                                & N                                        \\
DRB120-barrier-orig-no.c                 & \phantom{00}2.09                                    & N                                     & N                                                                & N                                        \\
DRB121-reduction-orig-no.c               & \phantom{0}23.44                                   & N                                     & N                                                                & N                                        \\
DRB124-master-orig-yes.c                 & \phantom{00}1.17                                    & P                                     & P                                                                & P                                        \\
DRB125-single-orig-no.c                  & \phantom{00}1.82                                    & N                                     & N                                                                & N                                        \\
DRB126-firstprivatesections-orig-no.c    & \phantom{00}1.06                                    & N                                     & N                                                                & N                                        \\
DRB139-worksharingcritical-orig-no.c     & \phantom{00}1.14                                    & N                                     & \cellcolor[HTML]{F4CCCC}P                                        & N                                        \\
DRB140-reduction-barrier-orig-yes.c      & \phantom{00}1.36                                    & P                                     & P                                                                & \cellcolor[HTML]{F4CCCC}N                \\
DRB141-reduction-barrier-orig-no.c       & \phantom{00}9.89                                    & N                                     & N                                                                & N                                        \\
DRB169-missingsyncwrite-orig-yes.c       & \phantom{00}2.80                                    & P                                     & P                                                                & P                                        \\
DRB170-nestedloops-orig-no.c             & \phantom{00}7.28                                    & N                                     & N                                                                & N                                        \\
DRB172-critical2-orig-no.c               & \phantom{00}9.09                                    & N                                     & N                                                                & N                                       \\ \hline
\end{tabular}
\end{adjustbox}
\vspace{0.1cm}
\caption{DataRaceBench results, part 2.}
\label{tbl:drbtwo}
\end{table}

\clearpage

\section{Change Log}

\subsection*{Version 2: 20 July 2023}

This version corrects an error in the previous version concerning
Definition \ref{def:detect}.  The old version called for a race check
when a thread arrives at an acquire state or departs from a state in
$R_i\setminus\LockingState_i$; when all threads are in the barrier;
and at a state with no enabled transition.  This does not suffice for
the correctness of Theorem \ref{thm:detect}.  A counterexample with
two threads is
\begin{align*}
  t_1 &\colon \code{x=1;\ } \lock(l)\code{;}\\
  t_2 &\colon \code{x=2;}.
\end{align*}
The one execution in the race-detecting state graph proceeds
\[
  x=1\ (t_1:\textit{check});\ x=2;\ \lock(l)\ (t1:\textit{clear});\ 
  (\textit{check-all}).
\]
This does not detect the race at the \textit{check-all} because $t_1$
cleared in the previous step.  The error in the proof of Theorem
\ref{thm:detect} occurs in Appendix \ref{sec:proofnobar}, where it is
assumed that the path $\beta$ is not empty.

This version changes Definition \ref{def:detect} so that a race check
occurs whenever a thread arrives at a state in $R_i$, a barrier state,
or a terminal state.  The check for races once all threads are in the
barrier is then redundant and has been removed.  The implementation
and its description in Section \ref{sec:transform} have been updated
accordingly.  The proof of Appendix \ref{sec:proofnobar} has been
corrected and is simpler.  Several other minor improvements were made
to CIVL, and the experiments were rerun.  The results are the same,
except for the times, which have been updated.  An updated link is
supplied for the experimental artifacts.

Various other minor changes and clarifications were made.

% Example 2:
% thread 1: x=1;
% thread 2: x=2;
% thread 3: while (true) {acq(l); rel(l);}
% Old way:
% x=1; x=2; acq(l); rel(l) (check for races with thread 3);
% There is no terminal state, so races involving threads 1,2 never get checked.

\subsection*{Version 1: 20 May 2023}

Original submission.